\newcommand{\fullver}[2]{{{\ifx\submissionversion\undefined#1\else#2\fi}}}

\documentclass[sigconf, nonacm, screen]{acmart}

\AtBeginDocument{%
  \providecommand\BibTeX{{%
    \normalfont B\kern-0.5em{\scshape i\kern-0.25em b}\kern-0.8em\TeX}}}

\usepackage{amsmath,amssymb,amsfonts}
\usepackage{graphicx}
\usepackage{textcomp}
\usepackage{xcolor}
\usepackage{ulem}

\usepackage{bm}
\usepackage{paralist}
\usepackage{booktabs}
\usepackage{amsthm}
\usepackage{color}
\usepackage{arydshln}
\usepackage{enumitem}
\usepackage{listings}
\usepackage{thm-restate}

\usepackage{hyperref}

\usepackage{xspace}
\usepackage{algorithmicx}
\usepackage{subcaption,graphicx,url,multirow}
\usepackage{algorithm}
\usepackage[noend]{algpseudocode}

\usepackage{comment}

\usepackage[switch]{lineno}
\usepackage[utf8]{inputenc}

\DeclareMathOperator{\argmax}{argmax}

\definecolor{ao}{rgb}{0.0, 0.5, 0.0}

\renewcommand{\emph}[1]{\textit{#1}}

\newcommand{\julian}[1]{{\color{blue} {\bf Julian:} #1}}
\newcommand{\changwan}[1]{{\color{purple} {\bf Changwan:} #1}}

\newcommand{\Ratio}[1]{Time} 
\newcommand{\avg}[1]{on average} 
\newcommand{\Avg}[1]{On average} 
\newcommand{\CONV}[1]{\ensuremath{\mathsf{#1}}}

\newcommand{\laxman}[1]{}
\newcommand{\cw}[1]{}
\newcommand{\cwcw}[1]{}
\newcommand{\cwappend}[1]{}

\newcommand{\R}[1]{}
\newcommand{\lr}[1]{}

\let\G\relax

\newcommand{\g}[1]{{\color{mygreen} #1}}
\newcommand{\G}[1]{{\color{mygreen} \bf #1}}

\newcommand{\stergiou}{\ensuremath{\mathsf{Stergiou}}}

\newcommand{\framework}{\ensuremath{\mathsf{GConn}}}
\newcommand{\connectit}{\ensuremath{\mathsf{ConnectIt}}}

\newcommand{\defn}[1]{\emph{\textbf{#1}}}

\newcommand{\myparagraph}[1]{\vspace{1.5pt}\noindent {\bf #1.}}
\newcommand{\id}[1]{\ifmmode\mathit{#1}\else\textit{#1}\fi}
\newcommand{\const}[1]{\ifmmode\mbox{\textc{#1}}\else\textsc{#1}\fi}

\newcommand{\makeset}{\textsc{MakeSet}}
\newcommand{\union}{\textsc{Union}}
\newcommand{\find}{\textsc{Find}}

\newcommand{\STAB}[1]{\begin{tabular}{@{}c@{}}#1\end{tabular}}

\def\BibTeX{{\rm B\kern-.05em{\sc i\kern-.025em b}\kern-.08em
    T\kern-.1667em\lower.7ex\hbox{E}\kern-.125emX}}

\newcommand{\koutsample}{$k$-out sampling}
\newcommand{\bfssample}{BFS sampling}

\newcommand{\hbsample}{HB sampling}

\definecolor{mygreen}{rgb}{0.0, 0.5, 0.0}

\newcommand{\mpc}{\ensuremath{\mathsf{MPC}}}

\newcommand{\connect}{\ensuremath{\mathsf{Connect}}}
\newcommand{\parentconnect}{\ensuremath{\mathsf{ParentConnect}}}
\newcommand{\extendedconnect}{\ensuremath{\mathsf{ExtendedConnect}}}
\newcommand{\rootupdate}{\ensuremath{\mathsf{RootUpdate}}}
\newcommand{\update}{\ensuremath{\mathsf{Update}}}
\newcommand{\shortcut}{\ensuremath{\mathsf{Shortcut}}}
\newcommand{\alter}{\ensuremath{\mathsf{Alter}}}
\newcommand{\fullshortcut}{\ensuremath{\mathsf{FullShortcut}}}

\mathchardef\mhyphen="2D
\newcommand{\unionasync}{\ensuremath{\mathsf{Union\mhyphen Async}}}
\newcommand{\unionhook}{\ensuremath{\mathsf{Union\mhyphen Hooks}}}
\newcommand{\unionearly}{\ensuremath{\mathsf{Union\mhyphen Early}}}
\newcommand{\unionremlock}{\ensuremath{\mathsf{Union\mhyphen Rem\mhyphen Lock}}}
\newcommand{\unionremcas}{\ensuremath{\mathsf{Union\mhyphen Rem\mhyphen CAS}}}
\newcommand{\unionremtsp}{\ensuremath{\mathsf{Union\mhyphen Rem\mhyphen TSP}}}
\newcommand{\ufmetaalgorithm}{\ensuremath{\mathsf{Connectivity (Union\mhyphen Find)}}}
\newcommand{\jayanti}{\ensuremath{\mathsf{Union\mhyphen JTB}}}
\newcommand{\liutarjan}{\ensuremath{\mathsf{Liu\mhyphen Tarjan}}}
\newcommand{\shiloachvishkin}{\ensuremath{\mathsf{Shiloach\mhyphen Vishkin}}}
\newcommand{\labelpropagation}{\ensuremath{\mathsf{LabelPropagation}}}

\newcommand{\eclcc}{\ensuremath{\mathsf{G\_ECL\mhyphen CC}}}
\newcommand{\afforest}{\ensuremath{\mathsf{G\_Afforest}}}

\newcommand{\ORIeclcc}{\ensuremath{\mathsf{ECL\mhyphen CC}}}
\newcommand{\gpucc}{\ensuremath{\mathsf{GPU\mhyphen CC}}}
\newcommand{\ORIafforest}{\ensuremath{\mathsf{Afforest}}}
\newcommand{\gswitch}{\ensuremath{\mathsf{GSWITCH}}}

\newcommand{\splitone}{\ensuremath{\mathsf{AtomicSplitOne}}}
\newcommand{\halveone}{\ensuremath{\mathsf{AtomicHalveOne}}}
\newcommand{\splice}{\ensuremath{\mathsf{SpliceAtomic}}}
\newcommand{\findnaive}{\ensuremath{\mathsf{Naive}}}
\newcommand{\findcompress}{\ensuremath{\mathsf{Compress}}}
\newcommand{\findhalve}{\ensuremath{\mathsf{AtomicHalve}}}
\newcommand{\findsplit}{\ensuremath{\mathsf{AtomicSplit}}}
\newcommand{\twotrysplit}{\ensuremath{\mathsf{TwoTrySplit}}}

\newcommand{\cas}{CAS}

\newcommand{\writemin}{writeMin}

\newcommand{\codevar}[1]{\mathit{#1}}

\begin{document}
\fancyhead{}

\title{Exploring the Design Space of Static and Incremental Graph Connectivity Algorithms on GPUs}
\titlenote{This is the full version of the paper appearing in Proceedings of the 2020 International Conference on Parallel Architectures and Compilation Techniques.}

\author{Changwan Hong}
\affiliation{MIT CSAIL}
\email{changwan@mit.edu}

\author{Laxman Dhulipala}
\affiliation{CMU}
\email{ldhulipa@andrew.cmu.edu}

\author{Julian Shun}
\affiliation{MIT CSAIL}
\email{jshun@mit.edu}

\begin{abstract}
Connected components and spanning forest are fundamental graph
algorithms due to their use in many important applications, such as
graph clustering and image segmentation. GPUs are an ideal platform
for graph algorithms due to their high peak performance and memory
bandwidth. While there exist several GPU connectivity algorithms in
the literature, many design choices have not yet been explored. In
this paper, we explore various design choices in GPU connectivity
algorithms, including sampling, linking, and tree compression, for
both the static as well as the incremental setting. Our various design
choices lead to over 300 new GPU implementations of connectivity, many
of which outperform state-of-the-art. We present an experimental
evaluation, and show that we achieve an average speedup of 2.47x
speedup over existing static algorithms. In the incremental setting,
we achieve a throughput of up to 48.23 billion edges per
second. Compared to state-of-the-art CPU implementations on a 72-core machine, we achieve a
speedup of 8.26--14.51x for static connectivity and 1.85--13.36x for incremental connectivity using a Tesla V100 GPU.


%
%

\end{abstract}



\maketitle

\section{Introduction}\label{sec:intro}

Connected components (connectivity) is a fundamental graph problem that plays a critical role in many graph applications.
Given an undirected graph with $n$ vertices and $m$ edges, the problem
assigns each vertex a label such that vertices that are reachable from each other have the same label, and otherwise have different labels~\cite{CLRS}.
Connectivity algorithms are used in many applications, such as computer vision~\cite{elhamifar2013sparse,ho2003clustering}, VLSI design~\cite{li2006connectivity},
and social analysis~\cite{haythornthwaite2005social}.
Graph connectivity is also a key subroutine to solve other graph algorithms, such as biconnectivity~\cite{tarjan1985efficient} and clustering~\cite{Ester1996,Patwary12,wen2017efficient,xu2007scan}, and some of these algorithms require many calls to graph connectivity.
As such, there has been a large amount of work on efficient parallel connectivity algorithms~\cite{Awerbuch1987,Chin1982,Chong95,Cole1991,Han1990,Hirschberg1979,Iwama1994,Johnson,Karger1999,Koubek85,Kruskal90,NathM82,Phillips89,Reif85,ShiloachV82,Gazit1991,Vishkin1984,SDB14,DhBlSh18,Greiner94,HalperinZ94,Slota14,BFGS12,Bader2005a,BaderJ96,MadduriB09,Bader2005b,Bus01,Caceres04,PatwaryRM12,Hambrusch,Hsu97,nguyen13,Hawick2010,Soman,Banerjee2011,ShunB2013,Sutton2018,LiuT19,Stergiou2018,Jain2017,Kiveris2014,Iverson2015,Chitnis2013,andoni2018parallel,behnezhad2019massively,behnezhad2019near}.

Graphics processing units (GPUs) are attractive devices for performing
graph computations because of their high computing
power and memory bandwidth. However, achieving high performance using
GPUs is challenging due to several factors, including uncoalesced memory access, insufficient parallelism to tolerate
high memory latency, load imbalance, and thread
divergence.
Several GPU connectivity implementations have been proposed in the
literature~\cite{Pai2016,Ben-Nun2017,Wang2016,Meng2019,Cong2014,Jaiganesh2018,Sutton2018},
but we found that there are many algorithmic choices and optimizations that have not been thoroughly explored for GPU connectivity.
The goal of this paper is to explore this large design space to better understand how different choices affect performance.

In this paper, we study \emph{min-based connectivity algorithms}~\cite{Jayanti2016,JayantiTB19,PatwaryRM12,LiuT19,Stergiou2018,ShunB2013}, which are based on vertices propagating labels to other connected vertices, and keeping the minimum label received. At convergence, vertices will have the same label if and only if they are in the same connected component.
A large number of our algorithms are based on using a union-find data structure for maintaining disjoint sets.  The data structure maintains
a tree for each sub-component found so far, and
joins trees to merge sub-components that are
connected.
We also study several other min-based algorithms that maintain trees, but not using a disjoint set data structure.
All existing GPU connectivity implementations are min-based; however,
there are
various design choices that have not been explored, such as different
rules for searching and compressing union-find trees and for
propagating labels. Furthermore, these algorithms can be improved by
using sampling as a preprocessing step to remove vertices in a large
component from consideration, so that the remaining steps are more
efficient. Sutton et al.~\cite{Sutton2018} provide one instantiation
of sampling combined with a particular union-find algorithm. Inspired
by their work, we explore different sampling strategies in this paper,
and combine them with each of our algorithms
to sweep the search space.

To implement different connectivity algorithms, we designed the  \framework{} framework, which is an extension of the \connectit{} framework~\cite{CONNECTIT} for multicore CPUs.
However, achieving high performance for connectivity algorithms on GPUs requires significant effort beyond what is provided in \connectit{} for two reasons.
The first reason is that the programming models on CPUs and GPUs are different,
which required us to significantly rewrite the codebase.
The second reason is that the bottlenecks on CPUs and GPUs are different (e.g., on GPUs, performance can be easily degraded from uncoalesced memory accesses,
low parallelism, and heavy use of atomics), and this required us to apply GPU-specific optimizations to achieve high performance.
\framework{} contains several GPU-specific optimizations: edge reorganization, which is specific to our connectivity algorithms, and CSR coalescing and vertex gathering, which are commonly used in other graph algorithms.

\begin{figure}[!t]
        \centering
        \includegraphics[width=\linewidth]{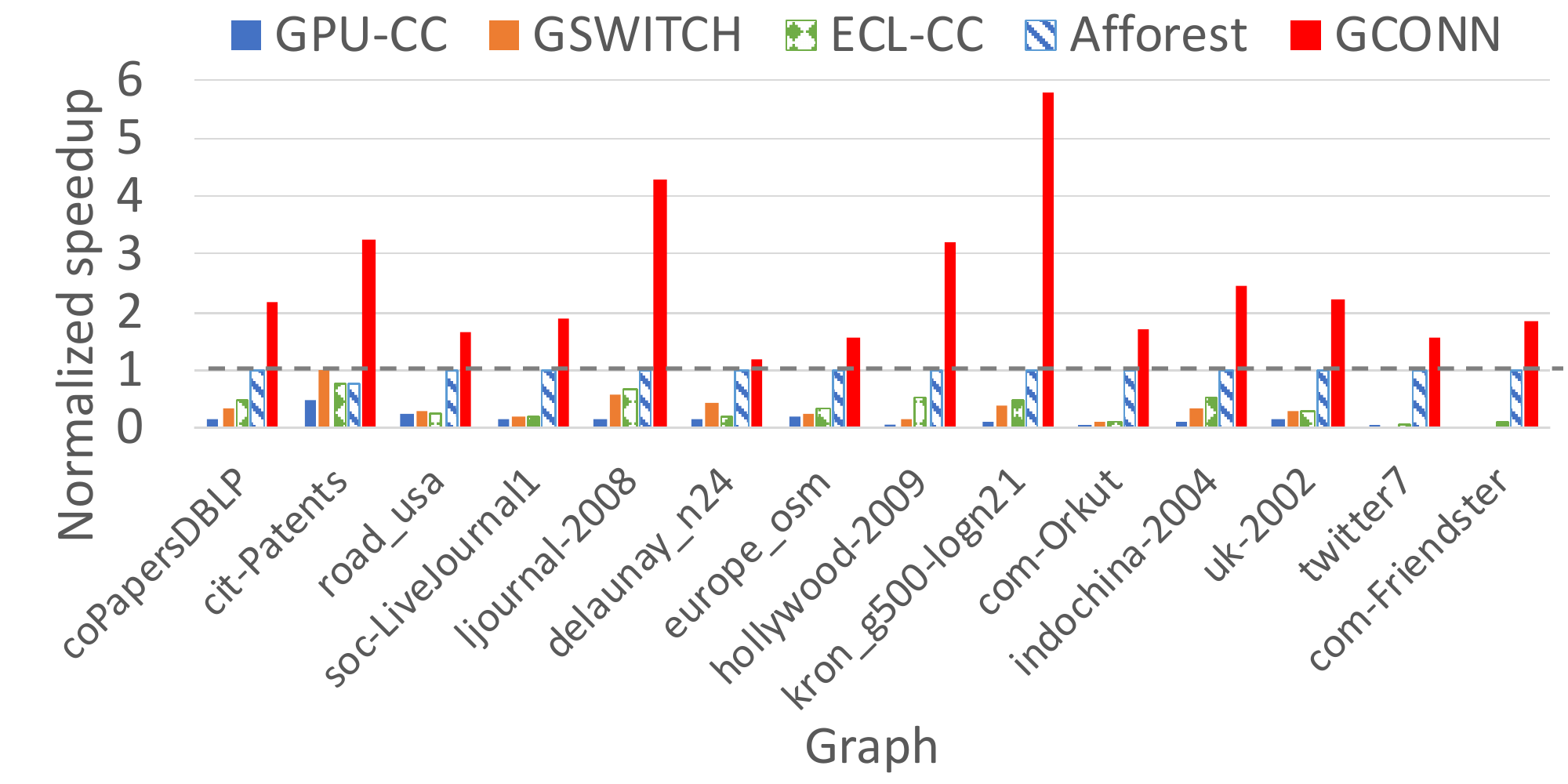}
        \vspace{-2em}
        \caption{Normalized speedup of four existing
          implementations/libraries and \framework{}. The fastest existing implementation's performance is normalized to 1.
        }
        \label{fig:intro}
\end{figure}

In this paper, we generate a total of 339 different connectivity
implementations and evaluate their performance.
With our comprehensive study, we are able to obtain the fastest GPU connectivity algorithms to date.
Figure~\ref{fig:intro} shows the normalized performance of the fastest implementation
in \framework{} compared to four state-of-the-art
GPU implementations: \CONV{GPU\mhyphen CC}~\cite{Soman},
\CONV{GSWITCH}~\cite{Meng2019}, \CONV{ECL\mhyphen CC}~\cite{Jaiganesh2018},
and \CONV{Afforest}~\cite{Sutton2018}.
We achieve an average speedup of 2.48x over the fastest implementation for each graph input.
Although we consider many connectivity implementations, we note that based on the results of our experimental study, practitioners only need to consider a handful of these implementations to obtain high performance.

In addition to connected components, most of our algorithms solve the
related problem of computing a spanning forest of a graph.
Furthermore, we extend our algorithms to the incremental setting,
where the connected components or spanning forest is updated upon new
edge arrivals. Our incremental algorithms are able to achieve
throughputs of up to 48.23 billion edges per second, which improves upon state-of-the-art for GPUs---\CONV{EvoGraph}~\cite{Sengupta17}---by orders of magnitude
based on a rough comparison of reported numbers since their code is not available.
Additionally, we compare our GPU implementations to the CPU implementations in \connectit{} on a 72-core CPU and show that we achieve speedups of 8.27--17.48x for static connectivity, and 1.85--13.36x for incremental connectivity on a Tesla V100 GPU that costs about half as much as the CPU.
Finally, we perform an analysis of different design choices to explain where the performance benefits of our fastest implementations are coming from.
 \framework{} is publicly available at
\url{https://github.com/hochawa/gconn}.

\myparagraph{Summary of Contributions}
We believe that this paper presents the most comprehensive study of GPU implementations of connectivity for both static and incremental connectivity to date.
By performing this extensive study, we provide an understanding of how different algorithmic choices affect performance
and where the performance benefits of fast connectivity implementations come from. Our paper provides the fastest GPU implementations of connectivity, which we obtain by combining many combinations of algorithmic choices that prior work did not explore. Building high-performance implementations of these optimizations and combining them is key to achieving high performance.
We believe that our proposed techniques can be integrated into graph processing compilers and other frameworks to significantly improve performance.

The remainder of the paper is organized as follows.
Section~\ref{sec:prelims} discusses graph terminology and prior work.
Section~\ref{sec:related} describes the design choices in \framework{},
and their relation to prior work.
We perform a comprehensive
evaluation of our implementations in Section~\ref{sec:eval},
and conclude in Section~\ref{sec:future}.

\section{Notation, Preliminaries, and Prior Work}\label{sec:prelims}

\myparagraph{Graph Notation and Formats} In this paper, we focus on
undirected, unweighted graphs, which we denote as $G=(V,E)$.
$V$ and $E$ are the set of vertices and edges, and $n=|V|$ and $m=|E|$ are the number of vertices and edges, respectively.
We consider two graph formats, coordinate list (COO) and compressed sparse row (CSR), with 0-based index notation (i.e., each vertex is labeled with a unique identifier in the range $[0,\ldots,n-1]$).
The COO format is represented using an array of edges, where each edge
contains a source and a destination vertex.
The CSR format contains two arrays, $\mathit{Offsets}$ and $\mathit{Edges}$, of lengths $n+1$ and $m$, respectively. The edges for vertex $i$ are stored in $\mathit{Offsets}[i],\ldots,\mathit{Offsets}[i+1]-1$, and $\mathit{Offsets}[n]=m$, and we assume they are in sorted order.

\myparagraph{Graph Connectivity Problems}
A \defn{connected component (CC)} in $G$ is a maximal set of vertices in which each pair of vertices in the component is connected by a path.
A connected components algorithm computes a vertex label $\mathit{labels}(v)$ for each $v \in V$. Two vertices $u,v \in V$ are in the same component if and only if $\mathit{labels}(u) = \mathit{labels}(v)$.
A \defn{connectivity query} returns true if and only if two input vertices belong to the same component.
A \defn{spanning forest (SF)} maintains a connected tree for each connected component.
A \defn{breadth-first search (BFS)} takes a graph with a source vertex and returns an array $A$ where $A(v)$ stores the shortest path between the source and $v$.
If a vertex $v$ is unreachable from the source vertex, $A(v)$ is set to infinity.

A \defn{union-find} (disjoint set) data structure keeps track of a number of disjoint sets such that
elements in each set have the same labels. Each set is represented by a tree and each element in the set has a parent pointer.
The parent of a tree root points to itself. In this paper, we assume that elements either point to themselves or point to a parent with a smaller ID.
The data structure provides three operations: \makeset{}, \union{}, and \find{}~\cite{CLRS}.
\makeset{($u$)} generates a new tree that is composed of only one node $u$, which is a root.
\union{($u,v$)} joins two trees, $u$ and $v$, into a single tree.
\find{($u$)} returns the label of the root of the tree containing $u$.
\find{} and \union{} operations can perform path compression during
execution to speed up subsequent operations. Union-find can be used
for connectivity by creating a set for each vertex, and calling
\union{} on the endpoints of each edge. The labels of the vertices can
be obtained in the end by running \find{} on each vertex.

In this paper, we also consider connectivity algorithms that maintain
trees, where we can move vertices between trees without completely
joining the trees (as needed for merging sets in union-find). We refer
to an algorithm as \defn{root-based} if it only modifies the parent
pointers of tree roots to point to vertices in other trees (vertices
are still free to update their pointer to point to other vertices in the
same tree). All of the union-find algorithms are root-based.

\myparagraph{Compare-and-Swap}
In this paper, we use the atomic \defn{compare-and-swap (CAS)} primitive, 
which takes as input a memory location, an old value, and a new value.
If the value stored in the memory location is equal to the old value, then the CAS atomically replaces the old value with the new value, 
and returns true. Otherwise, the CAS does not update the value, and returns false.

\myparagraph{Prior work on GPUs}
Soman et al.~\cite{Soman} provide {\bf GPU-CC}, the first high-performance implementation of the Shiloach-Vishkin algorithm~\cite{ShiloachV82} on GPUs.
Many libraries (e.g., \CONV{Gunrock}~\cite{Wang2017}, \CONV{IrGL}~\cite{Pai2016}, and \CONV{Groute}~\cite{Ben-Nun2017}) also adopt a variant of this approach.
Note that \gpucc{} uses the COO format so the performance does not suffer from poor load-balance on GPUs.
{\bf GSWITCH}~\cite{Meng2019} is a framework for graph processing which provides different combinations of optimization strategies for different graph algorithms, including connectivity.
Hence, \CONV{GSWITCH} allows tens of different combinations of optimizations
(e.g., different load-balancing strategies can be combined with other optimization strategies). The best set of optimizations is found using a machine learning approach.
{\bf ECL-CC}~\cite{Jaiganesh2018}
uses a concurrent union-find algorithm for connectivity in the CSR format.
{\bf Afforest}~\cite{Sutton2018} incorporates the \koutsample{} strategy (described in Section~\ref{sec:sampling}) that significantly improves performance on many real-world graphs.
\CONV{Afforest} uses the CSR format, and
uses the load-balancing technique from~\cite{Ben-Nun2017}.

\section{\framework{} Overview}\label{sec:related}

We first describe the overview of \connectit{} in
Section~\ref{sec:connectit}. Then we provide an overview of the
\framework{} framework in Section~\ref{sec:gconn}. In
Sections~\ref{sec:sampling} and~\ref{sec:finish}, we provide an
overview of specific algorithmic choices that we explore.
\fullver{
}{
Due to space constraints, we defer the complete description and pseudocode for some
of these methods to the full version of our paper~\cite{gconn_tech}.
}

\subsection{Overview of \connectit} \label{sec:connectit}
\connectit{} is a framework for multicore connectivity algorithms that outperforms existing state-of-the-art multicore algorithms
and generates various implementations for connectivity based on a two-phase execution model: a \emph{sampling} phase and a \emph{finish} phase.
In the sampling phase, a subset of edges are inspected to partially form connected components.
Next, the most frequently occurring label ($L_{\max}$) is identified.
In the finish phase,
only vertices whose label is not equal to $L_{\max}$ need to process their outgoing edges. Vertices with label $L_{\max}$ skip processing their edges, since any neighbor with label $L_{\max}$ is already in the same component, and any neighbor with label other than $L_{\max}$ will process an edge to this vertex.
This two-phase execution can significantly reduce the number of edges processed.
\connectit{} provides correctness proofs for the two-phase execution.
\framework{} is an extension of \connectit{} that we developed for
GPUs, which enables easy exploration of different algorithmic choices.

\subsection{\framework{} Framework}\label{sec:gconn}
In this section, we give an overview of the \framework{} that we use to
obtain our GPU implementations for static connectivity, spanning
forest, and incremental connectivity.

\begin{algorithm}[!t]
\caption{\framework{} for Static Connectivity} \label{alg:framework}
\footnotesize
\begin{algorithmic}[1]
\Procedure{StaticConn}{$G(V, E), \mathit{sample}, \mathit{finish}, \mathit{compress}$}
  \State $\mathit{labels} \gets \textproc{InitLabel}(V)$ \label{line:init}
  \State $\mathit{labels} \gets \textproc{SamplePhase}(G, \mathit{labels}, \mathit{sample}, \mathit{finish}, \mathit{compress})$ \label{line:sample}
  \State $L_{\max} \gets \textproc{GetMostFrequentLabel}(\mathit{labels})$ \label{line:getmax}
  \State $\mathit{labels} \gets \textproc{FinishPhase}(G, \mathit{labels}, \mathit{finish}, \mathit{comp})$ \label{line:finish}
  \State $\mathit{labels} \gets \textproc{LabelFinalization}(V, \mathit{labels})$ \label{line:finalization}
  \State \algorithmicreturn{}  $\mathit{labels}$ \label{line:return}
\EndProcedure
\end{algorithmic}
\end{algorithm}

\myparagraph{Static Connectivity}
Algorithm~\ref{alg:framework} presents the main steps in \framework{}.
\framework{} takes in a graph in either CSR or COO format, as well as
parameters for the sampling algorithm (\emph{sample}), the finish algorithm (\emph{finish}), and the compression algorithm (\emph{compress}).
The label of each vertex is initialized to its own ID on Line~\ref{line:init},
and the sampling phase is performed on Line~\ref{line:sample}.
On Line~\ref{line:getmax}, the most frequently occurring label ($L_{\max}$) is identified from the result of the sampling phase.
On Line~\ref{line:finish}, \framework{} performs the finish phase, processing edges of vertices with label not equal to $L_{\max}$.
Finally, on Line~\ref{line:finalization}, we finalize the labels of each vertex by assigning each vertex the label of the root of its tree.
We use C++ templates and inlined functions to achieve high-performance implementations while keeping the \framework{} implementations high-level.
Our implementations modularize the routines for the sampling algorithm, finish algorithm, compression algorithm, the load-balancing strategy, and graph format,
making it easy to test different implementations and add new variants.

\begin{algorithm}[!t]
\caption{\framework{} for Spanning Forest} \label{alg:framework_span}
\footnotesize
\begin{algorithmic}[1]
\Procedure{SpanningForest}{$G(V, E), \mathit{sample}, \mathit{finish}, \mathit{compress}$}
  \State $\mathit{labels} \gets \textproc{InitLabel}(V)$ \label{span:init}
  \State $\mathit{edges} \gets \textproc{InitEdge}(V)$ \label{span:init_edge}
  \State $\mathit{(labels,edges)} \gets 
   \textproc{SamplePhase}(G, \mathit{labels}, \mathit{sample}, \mathit{finish}, \mathit{compress})$  \label{span:sample}
  \State $L_{\max} \gets \textproc{GetMostFrequentLabel}(\mathit{labels})$ \label{span:getmax}
  \State $\mathit{(labels,edges)} \gets \textproc{FinishPhase}(G, \mathit{labels}, \mathit{finish}, \mathit{comp})$ \label{span:finish}
  \State \algorithmicreturn{}  $\mathit{edges}$ \label{span:return}
\EndProcedure
\end{algorithmic}
\end{algorithm}

\myparagraph{Static Spanning Forest}
As shown in Algorithm~\ref{alg:framework_span}, implementations for spanning forest are similar to those for connectivity;
\framework{} supports different combinations of sampling and finish methods while generating correct spanning forest algorithms.
In conjunction with the label initialization (Line~\ref{span:init}), we also maintain the edges
in the spanning forest using an auxiliary array of size $n$
(Line~\ref{span:init_edge}).
The key idea behind our implementations of spanning forest is to assign each edge of the discovered spanning forest to a unique vertex
that is one of the endpoints of the edge. This special vertex is the root vertex in the union-find structure that updates its parent pointer to point to another vertex, so that it is no longer a root.
Hence, the root-based algorithms for connectivity (to be described in Section~\ref{sec:finish}) can be converted to compute spanning forests.
The sampling phase simply creates a subset of the edges for the spanning forest (Line~\ref{span:sample}),
while computing partially connected components and determining $L_{\max}$, as in static connectivity (Line~\ref{span:getmax}).
After that, using $L_{\max}$, the finish phase computes the rest of the edges for the spanning forest (Line~\ref{span:finish}).
The finalization step is not required for spanning forest since the edges of the spanning forest have already been generated in previous steps, which eliminates the overhead for the post-processing.
In our experiments, we found that spanning forest algorithms are 6\% faster on average than their static connectivity counterpart.

\myparagraph{Incremental Connectivity}
As many real-world graphs are being updated frequently,
many connectivity algorithms have been proposed for dynamic graphs~\cite{McColl13,Ediger12,Simsiri2017,Sengupta17,Acar2019,dhulipala2020parallel}.
Many of our algorithms are a natural fit for the \emph{incremental} setting, where edges are inserted but not deleted.
We designed \framework{} to support incremental connectivity (and spanning forest) algorithms that
receive batches of operations consisting of edge insertions and connectivity queries that can be executed in parallel.
We will describe how \framework{} supports incremental connectivity algorithms in Section~\ref{sec:increment}.

\subsection{Sampling Algorithms}\label{sec:sampling}

As done in \connectit~\cite{CONNECTIT}, we decompose connectivity
algorithms into the sampling and finish phases.  The sampling phase
traverses a subset of edges in the graph to update the labels of
vertices.  The sampling phase can reduce the number of edges inspected
in the finish phase, since in practice we expect that a large fraction
of the vertices will already be settled in the $L_{\max}$ component
after applying the sampling phase.  All connectivity algorithms in the
literature today, except for \ORIafforest{}~\cite{Sutton2018}, only support
a finish phase. We implement sampling for graphs in CSR format due to the ease of skipping over all edges for particular vertices (i.e., the ones with label $L_{\max}$ after sampling). Below we introduce the different sampling methods implemented in \framework. We discuss them in the context of connectivity, although they are used similarly in spanning forest.
Pseudocode for these methods can be found in
\fullver{the Appendix.}{the full version of our paper~\cite{gconn_tech}.}

\myparagraph{$k$-out Sampling} Given a parameter $k$, \defn{\koutsample{}} computes connected components
on a \emph{sampled graph} constructed by uniformly sampling $k$ edges
out of each vertex~\cite{holm2019kout}.
Sutton et al.~\cite{Sutton2018} use a type of \koutsample{} strategy where
the \emph{first} $k$ edges out of each vertex are used.
The vertices obtain labels as a result of running a parallel connected components algorithm on the sampled graph.
In practice, after applying \koutsample{}, many of the vertices in the largest connected component will have the same label, since many real-world graphs have a single massive
component that most vertices will be a part of in the sampled graph.
In \framework, we implemented several variants of \koutsample{} for
different values of $k$, and found that taking the first 2 edges, as
in~\cite{Sutton2018}, gave the best performance overall.
Taking the first 2 edges per vertex enabled us to label most
of the vertices in the largest component with the final label, and minimized the overall number of edge inspections during the sampling and finish phases.
Moreover, since in many real-world graphs, the edges for a vertex are sorted
and if all vertices choose their first two edges (neighbors), there are likely to be many shared neighbors,
which improves locality and increases the size of the large component found.
For running connectivity on the subgraph, we
use different variants of union-find (discussed in Section~\ref{sec:finish}).

\myparagraph{Hook-Based Search (HB) Sampling}
Inspired by \ORIeclcc{}~\cite{Jaiganesh2018}, we designed the hook-based
search (HB) sampling approach to potentially reduce the number of edge
traversals compared to \koutsample{}. This approach uses a union-find
structure to maintain vertex labels.  First, each vertex $v$ inspects
its smallest (i.e., first) neighbor $w$, and updates its label to
$\mathit{labels}(v)=\min(\mathit{labels}(v),\mathit{labels}(w))$. This step is efficient since there is no
contention. Second, for all vertices that are still roots (i.e.,
$\mathit{labels}(v)=v$), we inspect their first $N$ edges and apply the union
operation to these edges.  The goal of these steps is
to minimize the total number of roots after sampling, since fewer
roots mean that the graph is more connected.  This approach can
potentially require fewer edge traversals than \koutsample{} for $k\ge
2$ if there are few roots after the first step.

\myparagraph{Breadth-First Search (BFS) Sampling} In breadth-first search (BFS) sampling, we
run a BFS from a chosen source vertex, which discovers the
connected component of this source. If the graph has a
massive connected component (containing a large fraction
of the vertices), we have a high probability of finding the
largest connected component.
Connected component algorithms typically incur a large overhead in a concurrent setting, whereas BFS with idempotent operations can incur a smaller overhead~\cite{Merrill2012}. Furthermore, the performance of BFS can be improved using direction-optimizing to reduce unnecessary edge traversals for many real-world graphs~\cite{Beamer12}.
In our \bfssample{} strategy, we applied a BFS from a source vertex, which we chose by sampling a subset of vertices and using the vertex with the largest degree from the sample. This idea was used by Slota et al.~\cite{Slota14} to speed up label propagation.

\subsection{Finish Algorithms}\label{sec:finish}
Here we introduce the finish algorithms in \framework{}. Again, we discuss them in the context of connectivity, but they are applied similarly in spanning forest.
Our algorithms are min-based algorithms, where vertices maintain labels, propagate labels to other connected vertices, and keep the minimum label received. When the algorithms converge, vertices will have the same label if and only if they are in the same component.
Pseudocode for our algorithms can be found in
\fullver{the Appendix.}{the full version of our paper~\cite{gconn_tech}.}

\myparagraph{Union-Find}
Union-find algorithms are a special case of min-based algorithms that use a disjoint set data structure to maintain and propagate labels.
\framework{} includes a broad set of concurrent union-find algorithms
that are obtained by combining different union operations with
different path compression strategies, all of which are root-based
algorithms.  In particular, it contains concurrent GPU implementations
of union used in Rem's algorithm (\unionremlock{}~\cite{PatwaryRM12}),
randomized linking by index (\unionasync{}~\cite{Jayanti2016}), and
randomized linking by rank (\ensuremath{\mathsf{Union\mhyphen}}\ensuremath{\mathsf{JTB}} \cite{JayantiTB19}).  Note that
\unionasync{} and \jayanti{} are lock-free compare-and-swap (CAS)
implementations, whereas \unionremlock{} is a lock-based
implementation.  Spin-locks are used in \unionremlock{}, which can
significantly degrade parallelism on GPUs~\cite{eltantawy2018warp}, so
we also implemented a lock-free version using CAS (\unionremcas{}).
We also implement two variations of \unionasync{}: \unionearly{}, which
traverses the paths of the two inputs simultaneously and terminates
once a common node is reached~\cite{Jayanti2016}; and \unionhook{},
which performs CAS operations on an auxiliary \emph{hooks} array so
that writes to the $\mathit{labels}$ array are uncontended (on the other
hand, \unionasync{} directly performs CAS operations directly on
the $\mathit{labels}$ array). All of our union operations link from larger to smaller ID to
ensure that there are no cycles. \find{} is implemented by traversing to the root of
the tree of the input vertex.

In our implementations, path compression is done on-the-fly when
calling either \union{} or \find{}, and \framework{} includes five
options: no path compression, path-splitting, path-halving, full path
compression, and path-splicing.
We describe these operations in detail in \fullver{the Appendix.}{the full version of our paper~\cite{gconn_tech}.}
Each union operation can be
combined with a subset of these path compression rules.
\connectit{}\cite{CONNECTIT} proves which of the combinations are
valid, and \framework{} supports the valid combinations
from~\cite{CONNECTIT}.

All of the implementations are asynchronous, meaning that \union{} and
\find{} calls can be executed concurrently without
synchronization. Furthermore, all of the implementations, except for
\unionremlock{}, are wait-free.  As far as we know, none of the variants
of union-find above have been implemented for GPUs in the
literature. The only union-find variants that have been implemented on
GPUs are \ORIeclcc{}~\cite{Jaiganesh2018} and \ORIafforest{}~\cite{Sutton2018}.
\ORIeclcc{}~\cite{Jaiganesh2018} implements a union-find algorithm, which
uses the \unionasync{} rule, but has a separate path compression step to
fully compress the paths, which requires
synchronization. \ORIafforest{}~\cite{Sutton2018} implements a variant of
the classic \shiloachvishkin{} algorithm~\cite{ShiloachV82}.
Their algorithm also has a separate compression
step, and requires synchronization. Both \ORIeclcc{} and \ORIafforest{} are
root-based algorithms. We also implemented these algorithms using
\framework{}.

\myparagraph{Other Min-based Algorithms}
Besides the union-find algorithms previously described, we also
implement the following min-based algorithms in \framework{}:
Shiloach-Vishkin (SV) algorithms~\cite{Awerbuch1987}, Liu-Tarjan (LT)
algorithms~\cite{LiuT19}, Stergiou's algorithm~\cite{Stergiou2018},
and the Label Propagation (LP) algorithm~\cite{ShunB2013}.
These other min-based algorithms generalize the union-find algorithms
by allowing a vertex $v$ in tree $T$ to be moved to another tree $T'$,
without requiring that all vertices in $T$ be moved to $T'$.  The
algorithms still link from larger to smaller ID to prevent cycles.

The SV algorithm is a classical parallel connectivity algorithm, and
many variants of it have been proposed in the literature,
e.g.,~\cite{Awerbuch1987,Cong2014,Bader2005a,Greiner94,BeamerAP15,Soman,Meng2019,Wang2017,Pai2016,Ben-Nun2017,Zhang19fastsv}.
Besides \ORIafforest{}, all other GPU
implementations~\cite{Soman,Pai2016,Ben-Nun2017,Wang2017,Meng2019} are
based on a GPU implementation by Soman et al.~\cite{Soman}. Soman et
al.'s algorithm alternates between hooking from smaller to larger
ID and from larger to smaller ID (the idea was originally proposed by
Greiner~\cite{Greiner94}), but does not guarantee that only roots are
hooked.
We also implemented a root-based variant of SV, where on every round
each vertex uses an atomic minimum operation to update their
neighbors' labels, followed by full path compression via pointer
jumping.

The LT algorithms are generated by combining several simple rules
about how to update the $\mathit{labels}$ array by using edges to transfer
connectivity information.  The updates are done using atomic $\min$
operations.  In their paper~\cite{LiuT19}, only five algorithms are
considered, but \connectit{} supports 16 variants by considering more
combinations of rules (all are min-based, and 6 variants are
root-based as well).  Stergious's
algorithm~\cite{Stergiou2018} is very similar to one of the LT variants,
but it maintains two $\mathit{labels}$ arrays, one for the previous iteration and
one for the current iteration.

Many implementations and frameworks for connectivity adopt the LP
algorithm, including~\cite{Slota14,Nguyen2013,ShunB2013}.  In each
round of the LP algorithm, the labels corresponding to the endpoints
of each edge are compared, and if they are different, the larger label updates
itself to be equal to the smaller label.  The LP algorithm terminates when there
is a round in which no labels are updated.

\begin{figure}[!t]
        \centering
        \includegraphics[width=\linewidth]{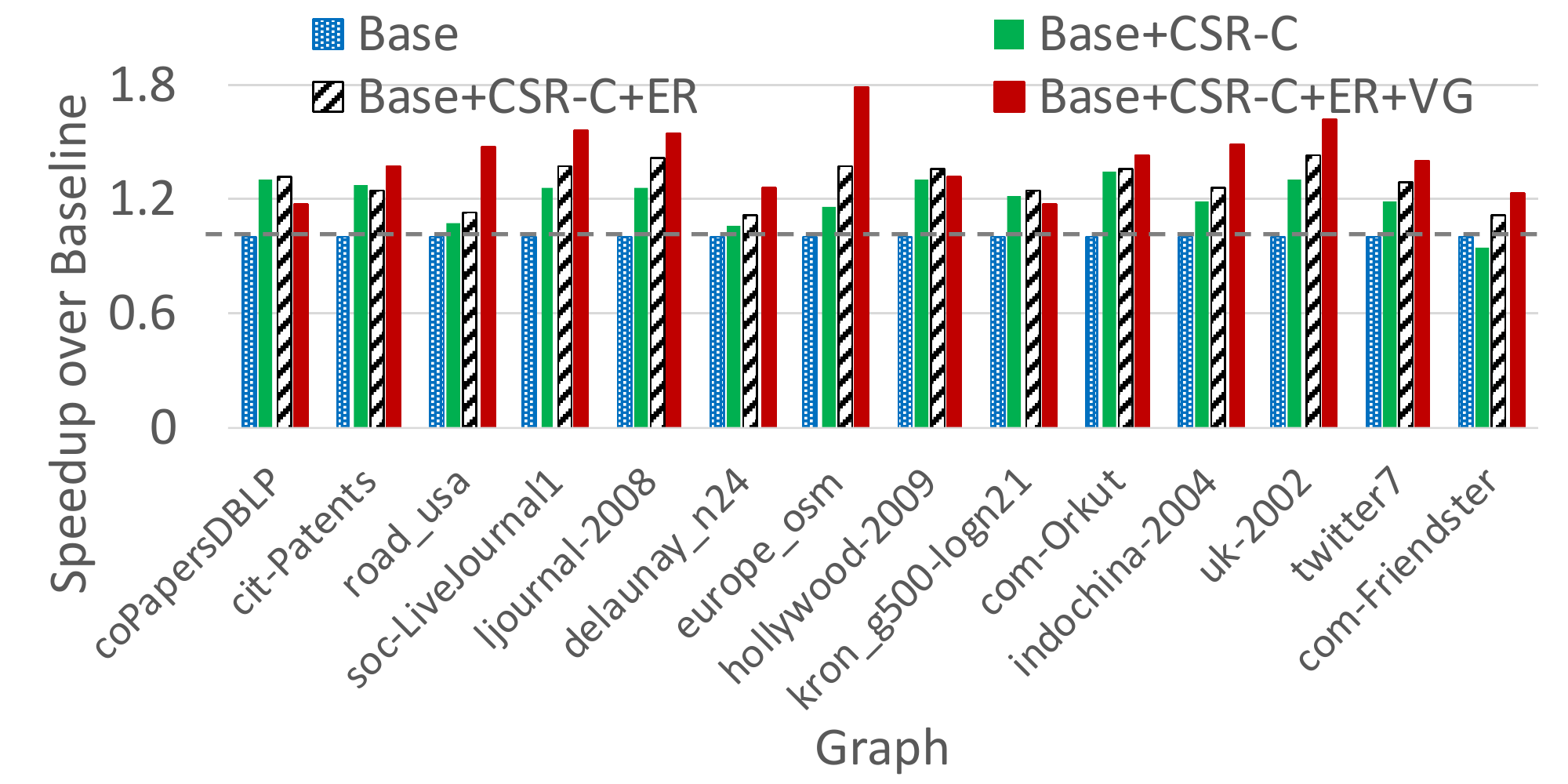}
        \vspace{-1em}
        \caption{The speedup of adding each optimization technique against Baseline for \unionasync{} with \emph{path compression} and \koutsample{}, which is the fastest connectivity variant on average. The performance of each graph is normalized to Baseline.}
        \label{fig:gpu_opt}
\end{figure}

\subsection{Iterative GPU Optimizations}
This section describes three
key GPU optimizations---\emph{CSR coalescing (CSR-C)}, \emph{edge
  reorganization (ER)}, and \emph{vertex gathering (VG)}---that we
apply to our implementations in \framework{}.  For practical purposes,
we only consider the optimization strategies that improve the
performance of the fastest \framework{} implementations.  Since we
found \ORIafforest{}~\cite{Sutton2018} to be the fastest existing
implementation in most cases, we identify the performance bottlenecks
in \ORIafforest{}, and show how to iteratively apply our optimizations
to achieve the performance of our fastest implementations. We first
substituted their union-find implementation with \unionasync{}, which
we found to be faster, and use this as the baseline.
Figure~\ref{fig:gpu_opt} shows how performance improves over the
baseline (\emph{Base}) with each additional optimization applied.

\myparagraph{CSR Coalescing (CSR-C)} We identified that the
sampling phase requires a significant amount of time in \ORIafforest{}
(35.15--98.63\%, with a median of 88.80\%), as shown in our experimental evaluation in
Section~\ref{sec:static_gpu_sample}.  During \koutsample{} with $k=2$,
\ORIafforest{} has two sub-phases, the first which traverses the first edge
out of every vertex, and the second which traverses the second edge
out of every vertex.  This strategy traverses the scattered entries in
the CSR array twice, which wastes memory bandwidth.  Since the first
two edges of a vertex are contiguous in the memory layout, the two
edges of a vertex can be processed simultaneously by two adjacent
threads, which makes memory accesses for the CSR array coalesced and
halves the data volume needed to access the CSR array. We observe that
applying this optimization on top of the baseline (\emph{Base+CSR-C}
in Figure~\ref{fig:gpu_opt}) improves performance by an average of
20.68\%.

\myparagraph{Edge Reorganization (ER)} During \koutsample{}, we found
that CAS operations make the warp-efficiency on the GPU very low due
to only a few threads in a warp being active.  For these algorithms,
if two edges having the same endpoint try to update a memory location
at the same time, then the updates can become serialized. This is
often the case in CSR format since edges incident to a single vertex
are processed in the same warp.
To alleviate this issue, we could have each thread read the first two
edges of a vertex and process them sequentially; however, this would
cause threads to have scattered accesses to the CSR edge array, which
decreases memory efficiency.
The edge reorganization optimization groups threads into pairs, where
each pair reads the two edges for one vertex together, and then reads
the two edges for another vertex together. Each time, the pair of
threads reads contiguous locations in memory. Then, using warp
shuffling~\cite{catanzaro2014decomposition,sung2014place} the four
edges can be reorganized so that the edges from the same vertex are
contiguous in memory, enabling each thread to process the two edges of
a vertex serially to avoid CAS operations.
\emph{Base+CSR-C+ER} in Figure~\ref{fig:gpu_opt} shows the performance
improvement after applying ER on top of CSR-C. We achieve an average
performance improvement of 28.79\% over the baseline.

\myparagraph{Vertex Gathering (VG)} The previous optimizations improve
the performance of the sampling phase. However, we found that the
performance of the finish phase can also be improved.  In the kernel
for the finish phase in \CONV{Afforest}, each thread accesses entries in the
$\mathit{labels}$ array in a cyclic fashion, and if the entry of the $i$'th
label is not equal to $L_{\max}$, then the edges of vertex $i$ are
processed in a load-balanced fashion by distributing the edges across
threads.  However, as we show in Section~\ref{sec:static_gpu_sample},
only a small subset of vertices have labels not equal to $L_{\max}$.
Therefore, many threads will be idle during the finish phase, which
degrades performance. To improve performance in the finish phase, we
first aggregate the vertices whose labels are not equal to $L_{\max}$
into an \emph{active vertex set}, and in the finish phase, each thread
reads a vertex in the active vertex set, and distributes the edges for
load-balancing.  We show the performance improvement of adding the
vertex gathering optimization in \emph{Base+CSR-C+ER+VG} of
Figure~\ref{fig:gpu_opt}. With all three optimizations, the overall
performance is 41.55\% faster than the baseline on average.

\subsection{Incremental Connectivity Support}\label{sec:increment}

\begin{algorithm}
\caption{\framework{} for Incremental Connectivity} \label{alg:framework_streaming}
\footnotesize
\begin{algorithmic}[1]

\Procedure{Incremental}{$G(V, E), \mathit{sample}, \mathit{finish}, \mathit{compress}$} \label{line:streaming_main}
\State $\mathit{labels} \gets \textproc{StaticConn}(G(V, E), \mathit{sample}, \mathit{finish}, \mathit{compress})$ \label{line:streaming_init}
\For{$B \in \mathit{Batches}$} \label{line:streaming_for}
\State $\mathit{(labels,query\_results)} \gets$ \par \label{line:streaming_process}
	  \hskip\algorithmicindent  $\textproc{FinishPhaseBatch}(B, \mathit{labels}, \mathit{finish}, \mathit{comp})$
\EndFor
\EndProcedure
\end{algorithmic}
\end{algorithm}

This section describes how \framework{} supports incremental
connectivity given batches of updates.  We support incremental
connectivity for the root-based algorithms, and
the pseudocode is shown in Algorithm~\ref{alg:framework_streaming}.
We first
generate the initial graph (which can be empty) by initializing the
$\mathit{labels}$ array for the vertices in the graph (Line~\ref{line:streaming_init}) using  the
\textproc{StaticConn} procedure from Algorithm~\ref{alg:framework}.
We assume that the $\mathit{labels}$ array is large enough to hold all of the vertices that will be encountered, but we only initialize the entries for vertices present in the initial graph.
Batches of insertions and connectivity queries arrive in an online
fashion, and are processed using one of the finish algorithms on just
the batch
(Lines~\ref{line:streaming_for}--\ref{line:streaming_process}).  The
batches are given in COO format, which is a natural input for
streaming algorithms. However, sampling is inefficient for graphs in
COO format, and hence we do not use sampling for incremental
algorithms.

Each batch $B$ is composed of a mix of inserts and queries. For an
update, \textproc{FinishPhaseBatch} updates the $\mathit{labels}$
array as \textproc{FinishPhase} from Algorithm~\ref{alg:framework}
does. Furthermore, a thread that inserts a new vertex will initialize
its entry in the $\mathit{labels}$ array by acquiring a spin lock.
For a query, \textproc{FinishPhaseBatch} calls the \find{} function
(one of the compression algorithms \framework{} provides) for both
query endpoints to check whether they are in the same component. The
query results ($\mathit{query\_results}$ in
Line~\ref{line:streaming_process}) are stored in a bitvector: the
$i$'th entry in the bitvector is true if the $i$'th edge of $B$ is a
query, and the two endpoints of it are in the same component.

As shown in
Algorithm~\ref{alg:framework_streaming}, the previous batches are
never inspected because the root-based algorithms guarantee
correctness \emph{without} requiring inspecting edges in a previous
batch~\cite{CONNECTIT}. The root-based algorithms incorporated into
\framework{} are all union-find algorithms used for static
connectivity, \shiloachvishkin{}, and the root-based \liutarjan{}
algorithms.  However, when \ensuremath{\mathsf{Union\mhyphen}}
\ensuremath{\mathsf{Rem\mhyphen CAS}} and \unionremlock{} with
\splice{} is used, the updates and queries need to be processed
separately to guarantee correctness~\cite{CONNECTIT}.

\section{Evaluation}\label{sec:eval}

In this section, we provide an experimental evaluation and analysis of
\framework{}.
All numbers reported in this section are the median of five runs on
the Volta machine unless noted otherwise.
We found that the trends for spanning forest are similar to the trends
for connectivity, with our spanning forest implementations obtaining
an average speedup of 6\% over the connectivity implementations due to
not requiring the label finalization step.

\myparagraph{Overview of Results}
The results of this section can be summarized as follows:

\begin{itemize}[topsep=0pt,itemsep=0pt,parsep=0pt,leftmargin=8pt]
	\item We provide an experimental evaluation of \framework{} connectivity implementations in the no sampling setting (Section~\ref{sec:static_gpu_no_sample}). 
	Without sampling, \unionasync{} and \unionremcas{} are the fastest implementations.
	\item In the sampling setting, we provide a detailed analysis of
  different sampling procedures and find that \koutsample{} or
  \hbsample{} can significantly improve the performance
 unless the average degree of vertices is low 
  (Section~\ref{sec:static_gpu_sample}).

	\item The fastest \framework{} algorithms consistently and significantly
		outperform state-of-the-art GPU connectivity implementations (Section~\ref{sec:state_of_the_art}).

	\item \framework{} incremental connectivity algorithms can achieve a throughput of tens of billions of edges per second.
		We also evaluate the throughput for different batch sizes, and ratios of insertions to queries (Section~\ref{sec:streaming_gpu}). 
		In the incremental setting, \unionasync{} is usually the fastest implementation.
		
	\item Compared to \connectit{}, a framework for CPU connectivity algorithms,
	\framework{} achieves 8.26--14.51x speedup for static connectivity and 1.85--13.36x speedup for incremental connectivity. 
    Our analyses also show GPUs are an attractive platform for
    connectivity algorithms in terms of both speed and cost
    (Section~\ref{sec:gpu_vs_gpu}).

\end{itemize}

\setlength{\tabcolsep}{2pt}
\begin{table}[!t]
  \centering
\footnotesize
\begin{tabular}{l|r|r|r|r|r}
\toprule
{\bf Dataset}                    & {\bf $n$} & {\bf $m$} & {\bf Diam.} & {\bf Num. Comps.} & {\bf Largest Comp.} \\
        \midrule
{coPapersDBLP}      & 540.49K   & 30.49M                      & 15*                       & 1                                 & 540.49K  \\ 
{cit-Patents}       & 3.77M     & 33.04M                        & 20*                       & 3,627                             & 3.76M  \\ 
{road\_usa}         & 23.95M    & 57.71M                      & 6,809                     & 1  					& 23.95M  \\ 
{soc-LiveJournal1}  	    & 4.85M     & 85.70M                      & 16                        & 1,876  				& 4.84M    \\ 
{ljournal-2008}     & 5.36M     & 99.03M                        & 31*                       & 75                                & 5.36M   \\ 
{delaunay\_n24}     & 16.78M    & 100.66M                     & 1,720*                    & 1                                 &  16.78M   \\ 
{europe\_osm}       & 50.91M    & 108.11M                     & 19,314*                   & 1                                 &  50.91M   \\ 
{hollywood-2009}    & 1.14M     & 112.75M                     & 11                        & 44,508                            &  1.07M    \\ 
{kron\_g500-logn21} 	    & 2.10M     & 182.08M                     & 6                         & 553,159                           &  1.54M  \\ 
{com-Orkut}         & 3.07M     & 234.37M                     & 9                         & 1   				& 3.07M   \\ 
{indochina-2004}    & 7.41M     & 301.97M                       & 26                        & 295                               &  7.32M   \\ 
{uk-2002}           & 18.52M    & 523.57M                     & 29*                       & 38,359                            &  18.46M   \\ 
{twitter7}           & 41.65M    & 2.41B                       & 23*                       & 1            &41.65M                       \\ 
{com-Friendster}    & 65.61M    & 3.61B                       & 32                        & 1            &65.61M                       \\ 
\end{tabular}
\caption{Graph inputs, including number of vertices ($n$), edges
($m$), diameter, number of connected components, and the largest
connected component. The graphs are symmetric, and edges are counted once in each direction.
For graphs on which we were unable to compute the exact diameter,
we compute the effective diameter (marked with *), which is a lower
bound on the actual diameter.
}
\label{table:sizes}
\end{table}

\myparagraph{Experimental Setup} \label{sec:exp_setup}
Our GPU evaluation is performed on two machines. The first is an NVIDIA Tesla V100, which is a Volta
generation GPU with 32\mbox{GB} that offers a 900
\mbox{GB/sec} memory bandwidth, 6\mbox{MB} of L2 cache, and 128\mbox{KB} of L1
cache per Streaming Multiprocessor (SM) with a total of 80 SMs.
The second is an NVIDIA TITAN Xp, which is a Pascal generation GPU with
12\mbox{GB} that offers a 547.6
\mbox{GB/sec} memory bandwidth, 3\mbox{MB} of L2 cache, and 48\mbox{KB} of L1
cache per SM with a total of 30 SMs.
All implementations are compiled with NVCC v10.0 using the \texttt{-O3} and \texttt{--use\_fast\_math} flags.

\myparagraph{Graph Data}
To show how \framework{} performs on various graphs of different
scales, we selected all publicly-available graphs used in \CONV{ECL\mhyphen CC}~\cite{Jaiganesh2018},
\CONV{EvoGraph}~\cite{Sengupta17}, and \connectit{}~\cite{CONNECTIT} that have more than 30M edges and fit in the GPU memory. 
Table~\ref{table:sizes} shows the details of our graph inputs,
including the number of vertices and edges, the graph diameter, the
number of connected components, and the size of the largest component.
Our inputs include many Web  and social network graphs that have
low diameters, as well as road networks that have high diameters.
All graphs that we use were obtained from SuiteSparse~\cite{Davis2011} or
SNAP~\cite{SNAP}. We symmetrized all of the graphs.

\begin{table*}[!t]
      \setlength{\tabcolsep}{1.5pt}
  \footnotesize
\centering
\begin{tabular}[t]{@{}c|  l | c | c | c | c | c | c | c | c | c | c | c | c | c | c}
  \toprule
  {\bf Group} & {\bf Algorithm} &  \multicolumn{1}{c|}{\bf \begin{tabular}[c]{@{}c@{}}coPapers\\DBLP\end{tabular}} & \multicolumn{1}{c|}{\bf \begin{tabular}[c]{@{}c@{}}cit-\\Patents\end{tabular}} 
& \multicolumn{1}{c|}{\bf road\_usa} & \multicolumn{1}{c|}{\bf \begin{tabular}[c]{@{}c@{}}soc-Live\\Journal1\end{tabular}}  &
\multicolumn{1}{c|}{\bf \begin{tabular}[c]{@{}c@{}}ljournal\\-2008\end{tabular}} & \multicolumn{1}{c|}{\bf  \begin{tabular}[c]{@{}c@{}}delaunay\\\_n24\end{tabular}} & \multicolumn{1}{c|}{\bf \begin{tabular}[c]{@{}c@{}}europe\\\_osm\end{tabular}} &
          \multicolumn{1}{c|}{\bf \begin{tabular}[c]{@{}c@{}}hollywood\\-2009\end{tabular}}  & \multicolumn{1}{c|}{\bf \begin{tabular}[c]{@{}c@{}}kron\_g500\\-logn21\end{tabular}}  &
                  \multicolumn{1}{c|}{\bf \begin{tabular}[c]{@{}c@{}}com-\\Orkut\end{tabular}} & \multicolumn{1}{c|}{\bf \begin{tabular}[c]{@{}c@{}}indochina\\-2004\end{tabular}} & \multicolumn{1}{c|}{\bf uk-2002}
& \multicolumn{1}{c|}{\bf twitter7} & \multicolumn{1}{c}{\bf \begin{tabular}[c]{@{}c@{}}com-\\Friendster\end{tabular}}\\ 

  \midrule
  \multirow{11}{*}{\STAB{\rotatebox[origin=c]{90}{No sampling}}}

   &	\unionearly{}&		1.49&		4.79&		5.68&		8.18&		8.83&		17.65&		6.87&		8.14&		16.23&		14.11&		26.39&		32.89&		375.50&		853.87 \\
 &	\unionhook{}&		0.68&		3.44&		5.56&		3.94&		5.26&		5.98&		11.67&		1.95&		4.65&		4.02&		7.36&		13.97&		155.98&		410.61 \\
 &	\unionasync{}&		0.46&		2.02&		\G{3.39}&	3.07&		\g{3.17}&	\g{3.84}&	6.76&		1.62&		3.64&		3.85&		\g{5.32}&	\g{9.09}&	\g{123.78}&	\g{365.98} \\
 &	\unionremcas{} &	\g{0.45}&	\g{1.88}&	3.83&		\g{2.98}&	3.20&		4.09&		\G{5.90}&	\g{1.53}&	\g{3.60}&	\g{3.70}&	5.39&		9.26&		133.46&		369.59 \\
   &	\unionremlock{}&	0.90&		4.66&		9.57&		5.73&		7.59&		10.31&		11.45&		2.73&		6.52&		4.19&		65.96&		93.69&		484.23&		487.72 \\
   &	\jayanti{}&		0.53&		2.84&		5.53&		3.70&		3.68&		5.29&		10.85&		1.96&		5.23&		5.78&		6.40&		11.07&		130.01&		400.17 \\
   &	\liutarjan{} &		1.20&		6.73&		12.13&		7.72&		11.27&		12.60&		31.23&		4.69&		12.39&		10.16&		21.53&		31.44&		464.76&		998.77 \\
   &	\shiloachvishkin{}&	 1.47&		13.05&		15.65&		12.23&		17.17&		18.63&		34.49&		7.57&		15.93&		34.09&		35.01&		47.73&		1056.17&	2066.00 \\
   &	\labelpropagation{}&	2.57&		8.98&		3289.59&	15.01&		58.89&		1032.09&	2.14e4&		6.18&		12.50&		18.58&		79.25&		112.74&		1361.37&	3646.96 \\
   &	\eclcc{}&		0.51&		2.26&		3.76&		3.18&		3.80&		4.19&		10.09&		1.92&		3.84&		5.01&		6.75&		11.58&		161.13&		385.48 \\
   &	\afforest{}&		5.96&		23.07&		4.01&		25.46&		42.12&		6.70&		19.07&		85.02&		91.66&		97.24&		43.12&		73.80&		143.75&		1661.75 \\

  \midrule
  \multirow{11}{*}{\STAB{\rotatebox[origin=c]{90}{\koutsample{}}}}

   &	\unionearly{}&		0.26&		3.30&		8.93&		1.28&		1.85&		6.34&		9.50&		0.74&		0.85&		1.00&		7.33&		7.83&		22.59&		49.34 \\
 &	\unionhook{}&		0.20&		2.52&		8.13&		1.05&		1.63&		4.87&		10.57&		0.39&		0.61&		0.74&		2.42&		4.39&		14.01&		31.09 \\
 &	\unionasync{}&		0.18&		\g{1.60}&	5.27&		\g{0.85}&	\G{1.19}&	3.60&		\g{6.95}&	0.35&		\G{0.51}&	0.63&		2.07&		\G{3.46}&	\g{10.91}&	27.79 \\
 &	\unionremcas{} &	\G{0.16}&	1.98&		5.98&		0.86&		1.38&		3.44&		7.78&		\G{0.33}&	0.53&		0.62&		\G{1.86}&	3.52&		12.81&		44.87 \\ 
   &	\unionremlock{}&	0.71&		3.79&		13.28&		1.80&		3.00&		8.46&		13.57&		0.98&		0.99&		0.74&		35.74&		41.10&		26.83&		53.61 \\
   &	\jayanti{}&		0.24&		2.31&		6.97&		1.27&		1.75&		4.66&		13.38&		0.47&		0.65&		0.86&		2.95&		5.04&		13.80&		32.20 \\
   &	\liutarjan{} &		1.29&		11.56&		20.71&		3.07&		13.96&		15.19&		40.17&		5.20&		13.05&		0.65&		20.20&		26.76&		41.69&		522.27 \\
   &	\shiloachvishkin{}&	1.55&		18.88&		21.35&		12.68&		17.37&		17.42&		47.61&		8.46&		15.99&		33.79&		32.47&		67.94&		1150.68&	2059.18 \\
   &	\labelpropagation{}&	2.28&		16.44&		8228.75&	2.67&		72.57&		895.65&		65136.83&	8.32&		18.65&		\g{0.60}&	103.27&		134.05&		27.01&		163.80 \\
   &	\eclcc{}&		0.19&		1.67&		\g{4.88}&	0.88&		1.28&		\G{3.19}&	7.28&		0.38&		0.56&		0.66&		2.11&		3.80&		11.14&		\g{27.29} \\
   &	\afforest{}&		0.24&		4.72&		5.63&		1.25&		4.40&		3.81&		9.58&		0.84&		2.41&		0.80&		2.79&		5.60&		13.46&		31.84 \\

  \midrule
  \multirow{11}{*}{\STAB{\rotatebox[origin=c]{90}{\hbsample{}}}}
   &	\unionearly{}&		0.41&		2.59&		17.79&		1.03&		1.86&		48.33&		25.42&		1.48&		0.76&		\G{0.49}&	12.95&		17.26&		12.62&		19.26 \\
 &	\unionhook{}&		0.24&		1.75&		12.45&		0.70&		1.41&		7.37&		16.78&		0.43&		0.64&		0.50&		3.00&		5.99&		9.90&		17.97 \\ 
 &	\unionasync{}&		0.24&		1.47&		8.60&		\G{0.67}&	\g{1.22}&	\g{5.95}&	14.99&		\g{0.41}&	0.57&		0.49&		2.60&		4.97&		9.12&		17.75 \\
 &	\unionremcas{} &	\g{0.24}&	1.56&		8.62&		0.67&		1.28&		6.37&		\g{13.66}&	0.41&		\g{0.57}&	0.50&		\g{2.43}&	\g{4.83}&	9.60&		18.30 \\
   &	\unionremlock{}&	1.51&		2.39&		14.37&		8.16&		1.90&		9.63&		17.11&		0.60&		1.59&		0.49&		10.60&		20.95&		11.78&		19.09 \\
   &	\jayanti{}&		0.70&		5.68&		8.54&		6.09&		5.37&		7.09&		18.86&		2.56&		11.11&		15.28&		9.78&		12.13&		339.15&		738.97 \\
   &	\liutarjan{} &		1.36&		11.68&		21.85&		1.91&		10.53&		24.67&		40.01&		3.70&		9.06&		0.61&		15.80&		27.15&		14.33&		353.80 \\
   &	\shiloachvishkin{}&	2.94&		18.55&		21.38&		6.33&		14.49&		24.37&		54.87&		9.20&		27.01&		0.64&		25.44&		67.48&		46.30&		829.49 \\
   &	\labelpropagation{}&	6.58&		16.62&		7943.94&	2.22&		73.01&		982.00&		6.65e04	&	8.24&		19.12&		0.57&		97.24&		125.14&		24.71&		123.77 \\
   &	\eclcc{}&		0.26&		\G{1.39}&	8.44&		0.71&		1.23&		6.32&		14.21&		0.47&		0.61&		0.50&		2.71&		5.28&		\G{8.78}&	\G{17.57} \\
   &	\afforest{}&		0.26&		2.85&		\g{8.04}&	0.73&		2.03&		9.12&		13.80&		0.55&		1.14&		0.50&		3.56&		6.22&		10.83&		17.75 \\

  \midrule
  \multirow{11}{*}{\STAB{\rotatebox[origin=c]{90}{\bfssample{}}}}
   &	\unionearly{}&		1.34&		2.39&		409.15&		2.10&		\g{2.67}&	90.27&		1778.75&	1.25&		0.81&		1.06&		80.15&		35.55&		22.20&		20.58 \\
 &	\unionhook{}&		1.35&		2.38&		391.73&		\g{1.90}&	2.68&		92.41&		1652.54&	\g{1.22}&	0.80&		1.03&		\g{79.23}&	\g{32.24}&	22.19&		20.56 \\
 &	\unionasync{}&		1.32&		2.40&		\g{390.92}&	1.93&		2.68&		92.91&		\g{1578.57}&	1.23&		0.83&		1.05&		80.58&		32.54&		22.22&		20.54 \\
 &	\unionremcas{} &	1.34&		\g{2.38}&	397.09&		2.07&		2.78&		\g{90.25}&	1598.75&	1.26&		0.82&		1.03&		79.64&		32.32&		\g{22.16}&	20.52 \\
   &	\unionremlock{}&	1.32&		2.38&		399.62&		2.06&		2.74&		92.41&		1584.45&	1.26&		\g{0.80}&	\g{1.02}&	83.84&		33.09&		22.19&		20.52 \\
   &	\jayanti{}&		1.36&		2.53&		406.20&		2.21&		3.06&		104.46&		1627.93&	1.29&		0.89&		1.18&		80.51&		32.89&		23.98&		23.72 \\
   &	\liutarjan{} &		1.38&		3.14&		395.90&		2.55&		2.92&		91.79&		1659.59&	1.44&		1.04&		1.10&		81.54&		35.14&		22.57&		21.19 \\
   &	\shiloachvishkin{}&	1.41&		3.07&		401.61&		2.50&		2.90&		97.07&		1616.75&	1.53&		1.05&		1.15&		82.49&		34.78&		23.29&		22.19 \\
   &	\labelpropagation{}&	1.39&		3.23&		403.12&		2.49&		2.77&		93.34&		1626.54&	1.51&		0.95&		1.07&		89.03&		40.99&		22.35&		20.86 \\
   &	\eclcc{}&		1.33&		2.41&		397.33&		2.06&		2.82&		94.21&		1631.24&	1.24&		0.83&		1.06&		81.36&		32.57&		22.20&		20.51 \\
   &	\afforest{}&		\g{1.30}&	2.39&		393.34&		2.06&		2.80&		91.45&		1625.83&	1.30&		0.83&		1.06&		83.19&		32.84&		22.18&		\g{20.48} \\

  \midrule
  \multirow{5}{*}{\STAB{\rotatebox[origin=c]{90}{\begin{tabular}[c]{@{}c@{}}Existing\\algorithms\end{tabular}}}}

&	\CONV{GPU\mhyphen CC}~\cite{Soman}&			2.77&	9.74&	22.85&	10.18& 	31.5&	22.83&	46.38&	21.38&	34.04&	21.25&	44.96&	49.01&	417.25&	$\mathsf{x}$\\
&	\CONV{GSWITCH}~\cite{Meng2019}&		1.07&	4.5&	21.03&	7.06&  	9.24&	8.22&	38.39&	6.82&	7.45&	7.54&	13.46&	26.1&	 $\mathsf{x}$&	$\mathsf{x}$\\
&	\CONV{ECL\mhyphen CC}~\cite{Jaiganesh2018}&		0.78&	5.89&	22.55&	7.14&  	7.71&	18.94&	25.97&	2.1&	6.02&	8.58&	9.06&	26.31&	258.56&	415.44\\
&	\CONV{Afforest}~\cite{Sutton2018}&		0.35&	6.13&	5.54&	1.27&  	5.07&	3.69&	9.13&	1.06&	2.97&	0.82&	4.55&	7.67&	13.5&	32.4				\\
   &   \CONV{BFS\mhyphen CC}&			 1.33&		721.59&		577.38&		277.35&		8.51&		89.33&		1749.63&	4267.22&	5.07e04&	0.96&		133.77&		3698.56&	21.07&		19.22 \\

\end{tabular}
\caption{Running times of implementations in \framework{} and
  state-of-the-art static connectivity algorithms in miliseconds on a V100 GPU.
We report running times for five groups: implementations with \emph{No sampling}, \emph{\koutsample{}}, \emph{\hbsample{}}, \emph{\bfssample{}}, and \emph{existing algorithms}. 
Within each of the first four group, we display the fastest variant for each graph in green.
For each graph, we also display the fastest variant across all groups in bold font.
$\mathsf{x}$ means that we were unable to obtain results due to the
graph not fitting in the GPU memory for the given implementation.  
}
  \label{tab:overall_gpu}
\end{table*}

\subsection{Static Parallel Connectivity without Sampling}\label{sec:static_gpu_no_sample}

In this section, we evaluate our \framework{} implementations for static connectivity algorithms in the \emph{No Sampling} setting.

\begin{figure}[!t]
        \centering
        \includegraphics[width=\linewidth]{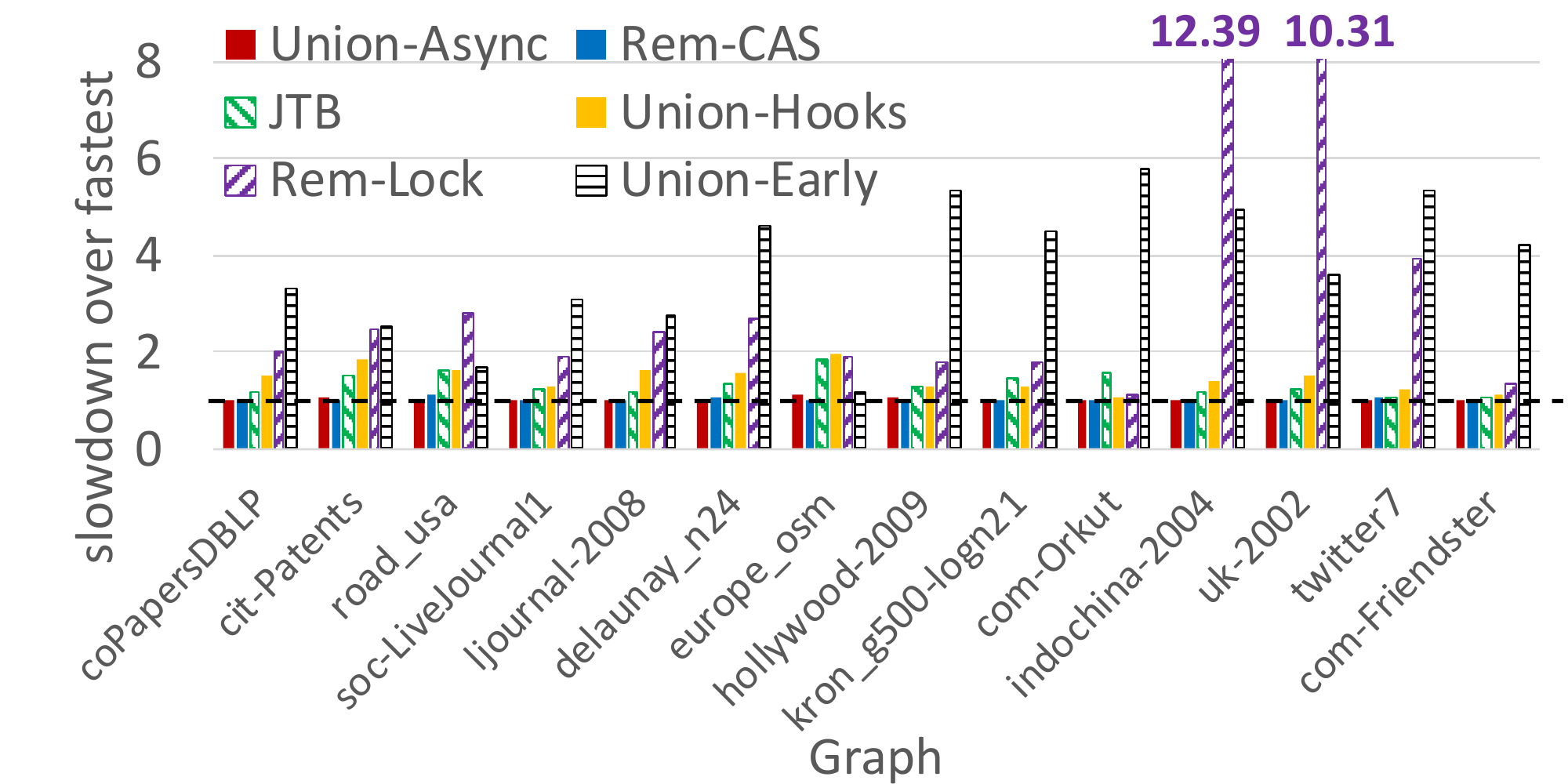}
        \vspace{-1em}
        \caption{The slowdown over the fastest union-find variants for each graph in the no-sampling setting.}
        \label{fig:uf_nosample}
\end{figure}
\begin{figure}[!t]
        \centering
        \includegraphics[width=\linewidth]{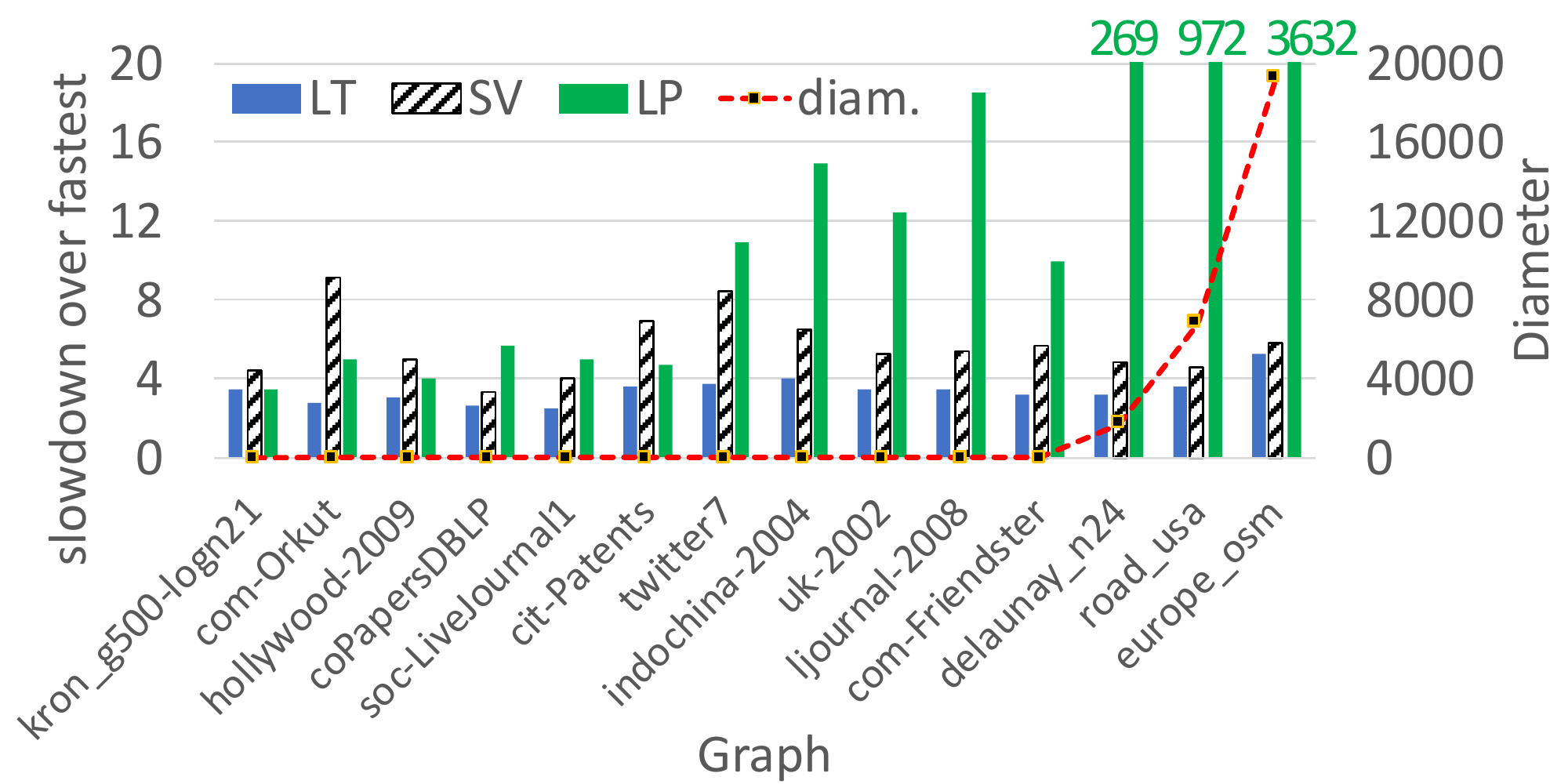}
        \vspace{-1em}
        \caption{Slowdowns for other min-based algorithms compared to the fastest union-find variants for each graph in the no-sampling setting. The graphs are sorted in ascending order of diameter.
        }
        \label{fig:min_nosample}
\end{figure}

\myparagraph{Evaluation of Union-Find Variants} The first group of rows in Table~\ref{tab:overall_gpu}
shows the results of the fastest implementations of each algorithm in the no sampling setting.

Figure~\ref{fig:uf_nosample} shows the slowdown of the fastest of each
of the union-find variants over the fastest overall variant for each graph.
For each graph, six bars are listed in order of average performance.
From Figure~\ref{fig:uf_nosample}, we observe that
the fastest implementation is either \unionasync{} or  \unionremcas{} in the \emph{No Sampling} setting.
\unionremcas{} is 1.02x slower \avg{} than \unionasync{} across all graphs,
and \jayanti{} is 1.26x slower \avg{} than \unionasync{} \R{A41} due to the usage of 64-bit CAS operations.
\ensuremath{\mathsf{Union\mhyphen}} \ensuremath{\mathsf{Hooks}} is designed to reduce the overhead for atomic operations, but it is 1.44x slower \avg{} than \unionasync{} due to a costly memory barrier needed to avoid race conditions.
\unionremlock{} is 3.81x slower \avg{} than \unionremcas{}
due to the poor performance of spin locks on GPUs.

\myparagraph{Evaluation of Other Min-based Algorithms}
Figure~\ref{fig:min_nosample} shows the slowdown of the other min-based algorithms compared to the fastest variant across all algorithms
in the no-sampling setting. The graphs that are sorted in order of
increasing diameter (presented with the red curve).

As shown in Figure~\ref{fig:min_nosample}, the other min-based algorithms are much slower than the union-find algorithms
because union-find algorithms only inspect each edge at once, whereas
the other algorithms typically traverse edges multiple times, which
leads to redundant computation.
The fastest \liutarjan{} and \shiloachvishkin{} variants are 3.72x and 5.19x slower \avg{} than \unionasync{} in the no-sampling setting, respectively.
We have included the \stergiou{} algorithm in the category of \liutarjan{} algorithms
as this algorithm is similar to \liutarjan{} algorithms; however, it is always much slower than the fastest variant of \liutarjan{} algorithms (up to 66x slower).

As shown for the three graphs located on the far right in Figure~\ref{fig:min_nosample},
the performance of \labelpropagation{} degrades significantly as the diameter increases.
This is because \labelpropagation{} needs a large number of rounds to propagate the minimum label to all vertices, and during that time most vertices are active.
Even on low-diameter graphs (diameter at most 32 in our experiments), \labelpropagation{} is still 7.76x slower \avg{} than the union-find algorithms since \labelpropagation{} still requires multiple traversals per edge, whereas the union-find algorithms do not.

As shown in Table~\ref{tab:overall_gpu}, we also incorporate the finish algorithms used in two fastest state-of-the-art works,
\ORIeclcc{}~\cite{Jaiganesh2018} (denoted as \eclcc{}) and \ORIafforest{}~\cite{Sutton2018} (denoted as \afforest{}), into \framework{},
and evaluate them in Section~\ref{sec:state_of_the_art}.

\subsection{Static Parallel Connectivity with Sampling}\label{sec:static_gpu_sample}

This section studies how our three sampling strategies affect performance, in terms of their execution time and the quality of the resulting sub-problem that they generate for the finish step in \framework{}. We start by studying how \koutsample{} performs.

\begin{figure}[!t]
        \centering
        \includegraphics[width=\linewidth]{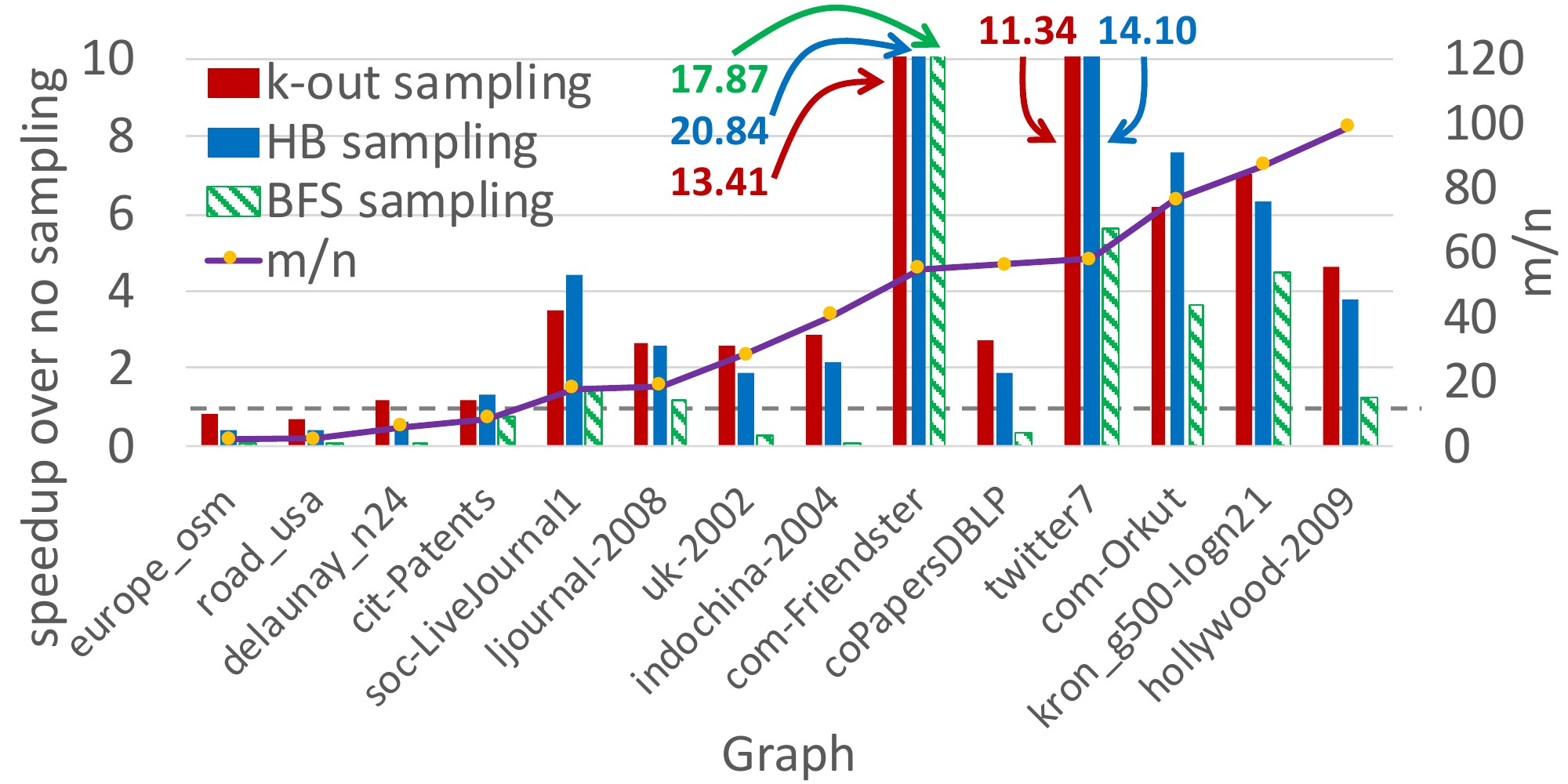}
        \vspace{-1em}
        \caption{The speedup of the best variant of each sampling algorithm over the best variant in the no-sampling setting.
	The graphs are sorted in ascending order of $m/n$ (i.e., the average degree).}
        \label{fig:sample_speedup}
\end{figure}

\myparagraph{Evaluation of \koutsample{}} The second group of rows in Table~\ref{tab:overall_gpu}
presents the results of the fastest implementations of each algorithm using \koutsample{}.
The fastest implementation is \unionasync{}, \unionremcas{}, or \eclcc{};
\unionremcas{} is 1.03x slower \avg{} than \unionasync{}, and
we defer the analysis for \eclcc{} to Section~\ref{sec:state_of_the_art}.

Figure~\ref{fig:sample_speedup} shows how each sampling algorithm improves performance.
To see the performance improvement with respect to the average degree of the vertices (i.e., $m/n$),
we sorted the graphs in ascending order of the average degree.

For the union-find variants, except on the two road-network graphs (\emph{road\_usa} and \emph{europe\_osm}),
\koutsample{} improves performance over the unsampled versions by 6.16x \avg{} due to the significant reduction of edges that need to be inspected, as we will show shortly.
For the other min-based algorithms, other than on the two road-network graphs, \koutsample{} improves performance by 4.61x \avg{} over the unsampled versions.

As shown in Figure~\ref{fig:sample_speedup}, when the average degree is small (e.g., \emph{road\_usa} and \emph{europe\_osm}),
\koutsample{} as well as other sampling algorithms degrades performance.
This is because most of the edges in the graph get inspected in the
sampling phase, and there is additional overhead to split the computation across two phases.
In general, \koutsample{} decreases the number of edge inspections by up to $m - kn$, and so as
 the average degree increases, we can expect a higher potential performance improvement.
 As shown in Figure~\ref{fig:sample_speedup}, \koutsample{} usually shows better performance when the average degree is large.

\myparagraph{Evaluation of \hbsample{}} The third group of rows in Table~\ref{tab:overall_gpu}
presents the fastest variants of each algorithm using \hbsample{} with a default value of $N=4$, which we found to work the best on average across our input graphs.\footnote{
Although the optimal value of $N$ can vary based on the input graph, 
determining the optimal  value on a per-graph basis  would incur significant overhead,
which would outweigh the benefits of using the optimal parameter as opposed to the default parameter.}
Unlike \koutsample{}, \hbsample{} does not always improve performance; in many cases, the performance is degraded.
The overall trend of \hbsample{} is similar to that of \koutsample{}.
As shown in Figure~\ref{fig:sample_speedup}, when the average degree of vertices is small, \hbsample{} significantly degrades performance, and
otherwise \hbsample{} can greatly improve performance.
We will discuss how \hbsample{} compares with our other sampling schemes shortly.

\myparagraph{Evaluation of \bfssample{}} The fastest variants using \bfssample{} are presented in the fourth group in Table~\ref{tab:overall_gpu}.
The effectiveness of \bfssample{} is dependent on the diameter of an input graph and the parameters for the direction-optimizing BFS.
As shown in Figure~\ref{fig:sample_speedup}, for the high-diameter graphs (the three graphs on the far left),
\bfssample{} degrades performance by 24.59x \avg{} over the unsampled versions
due to the very small active vertex set for each BFS iteration and since a GPU kernel must be launched for each iteration.
For the other graphs, \bfssample{} achieves a 0.07--174.87x speedup over the unsampled versions.
Note that as shown in the last row of Table~\ref{tab:overall_gpu}, we also implemented the \CONV{BFS\mhyphen CC} algorithm introduced in \CONV{Ligra}~\cite{ShunB2013},
in which a BFS is repeated on each new component until all components are found.
\CONV{BFS\mhyphen CC} performs poorly due to the GPU kernel launch overhead for
high-diameter graphs and for graphs with many connected components.

\begin{table}[!t]
\footnotesize
\centering
\renewcommand{\tabcolsep}{0.5 mm}
\begin{tabular}[t]{l|c|c|c|c|c|c|c|c|c}
  \toprule

\multicolumn{1}{c}{{\bf Graph}} & \multicolumn{1}{c}{\begin{tabular}[c]{@{}c@{}}BFS\\ \Ratio{}\end{tabular}} & \multicolumn{1}{c}{\begin{tabular}[c]{@{}c@{}}BFS\\ Cov\end{tabular}} & \multicolumn{1}{c}{\begin{tabular}[c]{@{}c@{}}BFS\\ IC\end{tabular}} & \multicolumn{1}{c}{\begin{tabular}[c]{@{}c@{}}KOut\\ \Ratio{}\end{tabular}} & \multicolumn{1}{c}{\begin{tabular}[c]{@{}c@{}}KOut\\ Cov\end{tabular}} & \multicolumn{1}{c}{\begin{tabular}[c]{@{}c@{}}KOut\\ IC\end{tabular}} & \multicolumn{1}{c}{\begin{tabular}[c]{@{}c@{}}HB\\ \Ratio{}\end{tabular}} & \multicolumn{1}{c}{\begin{tabular}[c]{@{}c@{}}HB\\ Cov\end{tabular}} & \multicolumn{1}{c}{\begin{tabular}[c]{@{}c@{}}HB\\ IC\end{tabular}} \\
  \midrule
{coPapersDBLP}&		60.3\%&		100.0\%&	0.0\%&			46.7\%&	98.9\%&		0.2\%  &	   97.2\%&	89.1\% &	5.4\% \\	
{cit-Patents}&		83.9\%&		99.7\%&		0.0\%&			90.1\%&	98.2\%&		0.6\% 	&	   99.5\%&	96.0\% &	2.3\% \\
{road\_usa}&		86.8\%&		100.0\%&	0.0\%&			85.4\%&	95.9\%&		3.6\% 	&	   83.1\%&	0.0\% &	100.0\% \\
{soc-LiveJournal1}&		86.9\%&		99.9\%&		0.0\%&			91.0\%&	99.9\%&		0.0\%    &         92.1\%&	99.5\% &	1.1\% \\
{ljournal-2008}&	92.1\%&		100.0\%&	0.0\%&			83.7\%&	99.3\%&		0.2\% 	&	   92.0\%&	94.9\% &	2.5\% \\
{delaunay\_n24}&	92.3\%&		100.0\%&	0.0\%&			76.9\%&	100.0\%&	0.0\% &	  	   89.3\%&	0.0\% &	100.0\% \\
{europe\_osm}&		92.4\%&		100.0\%&	0.0\%&			79.2\%&	100.0\%&	0.0\% 	&	   92.4\%&	0.0\% &	100.0\% \\
{hollywood-2009}&	55.1\%&		93.8\%&		0.1\%&			42.0\%&	91.0\%&		0.5\% &	 	  99.8\%&	87.4\% &	2.0\% \\
{kron\_g500-logn21}&		93.5\%&		73.6\%&		0.0\%&			41.2\%&	73.6\%&		0.0\% 	&	   99.7\%&	73.5\% &	0.0\% \\
{com-Orkut}&		74.4\%&		100.0\%&	0.0\%&			54.4\%&	100.0\%&	0.0\% 	&	   97.8\%&	100.0\% &	0.0\% \\
{indochina-2004}&	90.3\%&		98.7\%&		1.3\%&			42.5\%&	86.8\%&		7.4\% 	&	   99.5\%&	63.2\% &	24.4\% \\
{uk-2002}&		95.9\%&		99.7\%&		0.1\%&			90.2\%&	92.0\%&		4.9\% 	&	   88.2\%&	72.4\% &	20.8\% \\
{twitter7}&		93.4\%&		100.0\%&	0.0\%&			39.4\%&	100.0\%&	0.0\% 	&	   99.7\%&	99.5\% &	0.0\% \\
{com-Friendster}&		98.0\%&		100.0\%&	0.0\%&			95.2\%&	100.0\%&	0.0\% 	&	   90.3\%&	99.9\% &	0.0\% \\	
  \bottomrule
\end{tabular}
\caption{ 
This table presents how effective the sampling strategy is for each of our graph inputs.
The {\bf \Ratio{}} columns show the percentage of sampling time to the total execution time.
The {\bf Cov} columns show the percentage of vertices that are in the largest connected component after the sampling phase.
Hence, the edges incident to these vertices are not inspected in the finish phase.
The {\bf IC} columns show the percentage of inter-component edges that will be processed in the finish phase.
}
\label{tab:sample_cover}
\end{table}

\begin{figure}[!t]
        \centering
        \includegraphics[width=\linewidth]{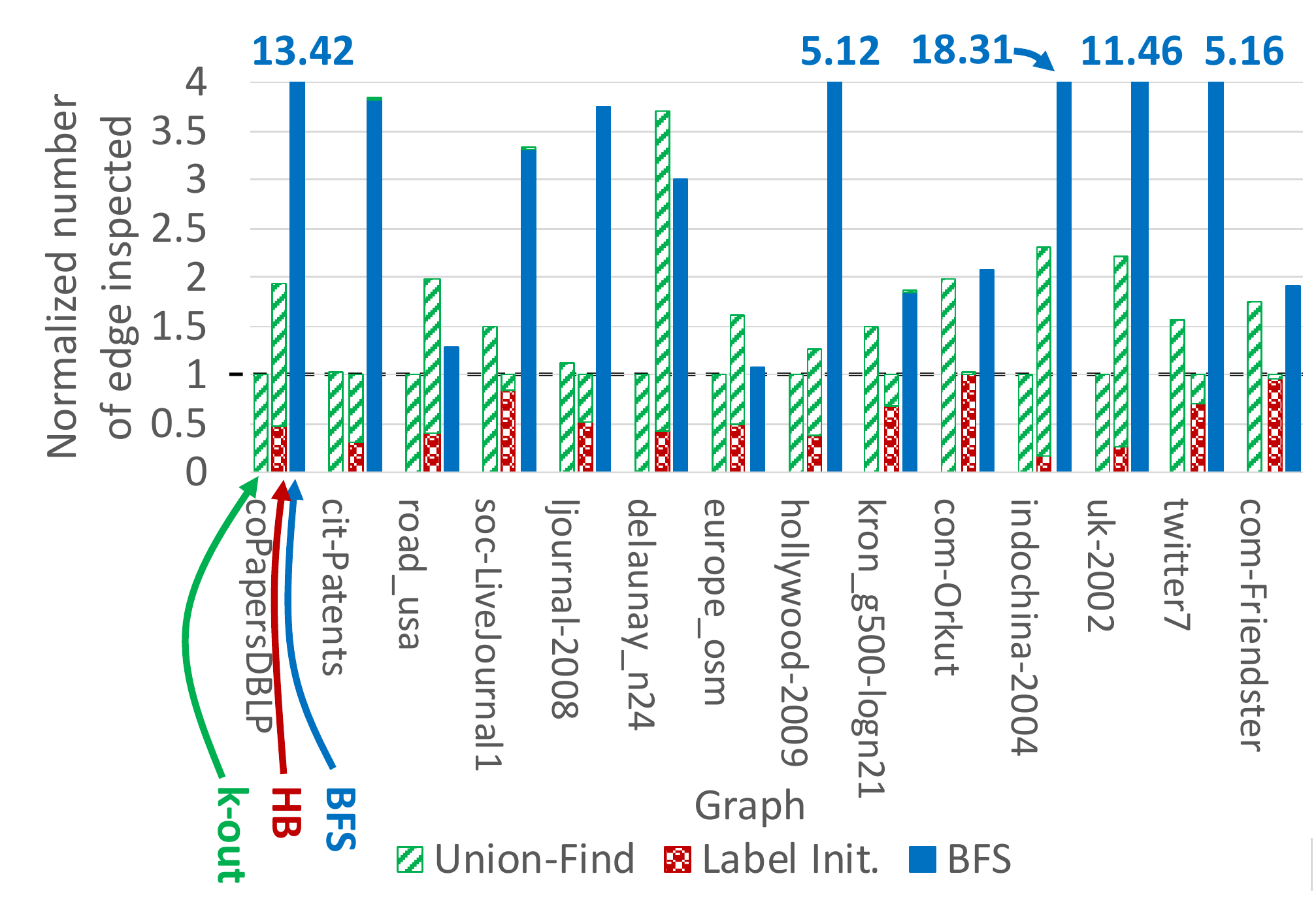}
        \vspace{-1em}
        \caption{The normalized number of edges inspected for each sampling strategy. The first bar for each graph (green only) corresponds to the number of edge inspections for $k$-out using union-find. The second bar for each graph (red and green) shows the number of edge inspections in the label initialization step, and the number of edges inspected by the union-find step of \hbsample{} respectively. Finally, the blue bar shows the number of edges inspected in \bfssample{}.}
        \label{fig:sample_comp}
\end{figure}

\myparagraph{Comparing Different Sampling Strategies}
Table~\ref{tab:sample_cover} shows how many vertices and edges are covered by each sampling algorithm.
As shown in Table~\ref{tab:sample_cover}, since BFS traverses all vertices connected to a source vertex likely to be in the largest connected component,
the fraction of vertices covered ({\bf Cov}) and the fraction of remaining inter-component edges ({\bf IC}) are maximized and minimized, respectively.
Unfortunately, as shown in Table~\ref{tab:overall_gpu}, \bfssample{} takes significantly longer than other sampling algorithms,
especially for the high-diameter graphs (\emph{delaunay\_n24}, \emph{road\_usa}, and \emph{europe\_osm})
due to the kernel launch overhead and low parallelism on these graphs. It also inspects significantly more edges than the other two schemes, as we will show shortly.
Although \bfssample{} usually takes longer than \koutsample{}, we observe that {\bf Cov} and {\bf IC} for \bfssample{} and \koutsample{} are very similar on our inputs.
The sampling time for \hbsample{} is much lower than for the others due to the reduced number of edge inspections, but for some graphs, {\bf CoV} and {\bf IC} are significantly lower than for the other two schemes (especially for high-diameter graphs).
The {\bf \Ratio{}} columns in Table~\ref{tab:sample_cover} show the ratio of the sampling time to the total execution time. We see that
the sampling phase takes most of the time,
and hence, the design of an efficient, high-quality sampling algorithm is of paramount importance.

Figure~\ref{fig:sample_comp} shows the normalized total number of edges inspected during the execution of each sampling algorithm.
For each graph, Figure~\ref{fig:sample_comp} shows three bars. The first bar (green only) corresponds to the number of edges inspected by \koutsample{} using union-find. The second bar (red and green) shows the number of edges inspected during the label initialization step, and the union-find step of \hbsample{}, respectively. Finally, the third bar (blue) shows the number of edges inspected by \bfssample{}.
From Table~\ref{tab:overall_gpu} and Figure~\ref{fig:sample_comp},
we observe that \bfssample{} is never the fastest, and \hbsample{} is the fastest
when the number of edge inspections for union-find  is small during the sampling phase.
In this case, the total number of edge inspections is also minimized.
For \emph{com-Friendster}, as shown in Table~\ref{tab:overall_gpu}, \bfssample{} is faster than \koutsample{};
the number of edge inspections with \bfssample{} is usually the highest, but for \emph{com-Friendster}, the total number of edge inspections for both \bfssample{} and \koutsample{}
is similar, and thus \bfssample{} is faster because
the edge inspection step performed by the BFS (a single CAS on one of the endpoints) is much faster than the edge inspection by a union-find algorithm (a loop that must potentially run multiple times due to contention).
As seen in Figures~\ref{fig:sample_speedup} and \ref{fig:sample_comp}, for some graphs (e.g., \emph{indochina-2004}, which has a speedup of 0.07x), the number of edge inspections is much higher for BFS, leading to significant performance degradation.

\subsection{Comparison with State-of-the-art}\label{sec:state_of_the_art}

\begin{figure}[!t]
        \centering
        \includegraphics[width=\linewidth]{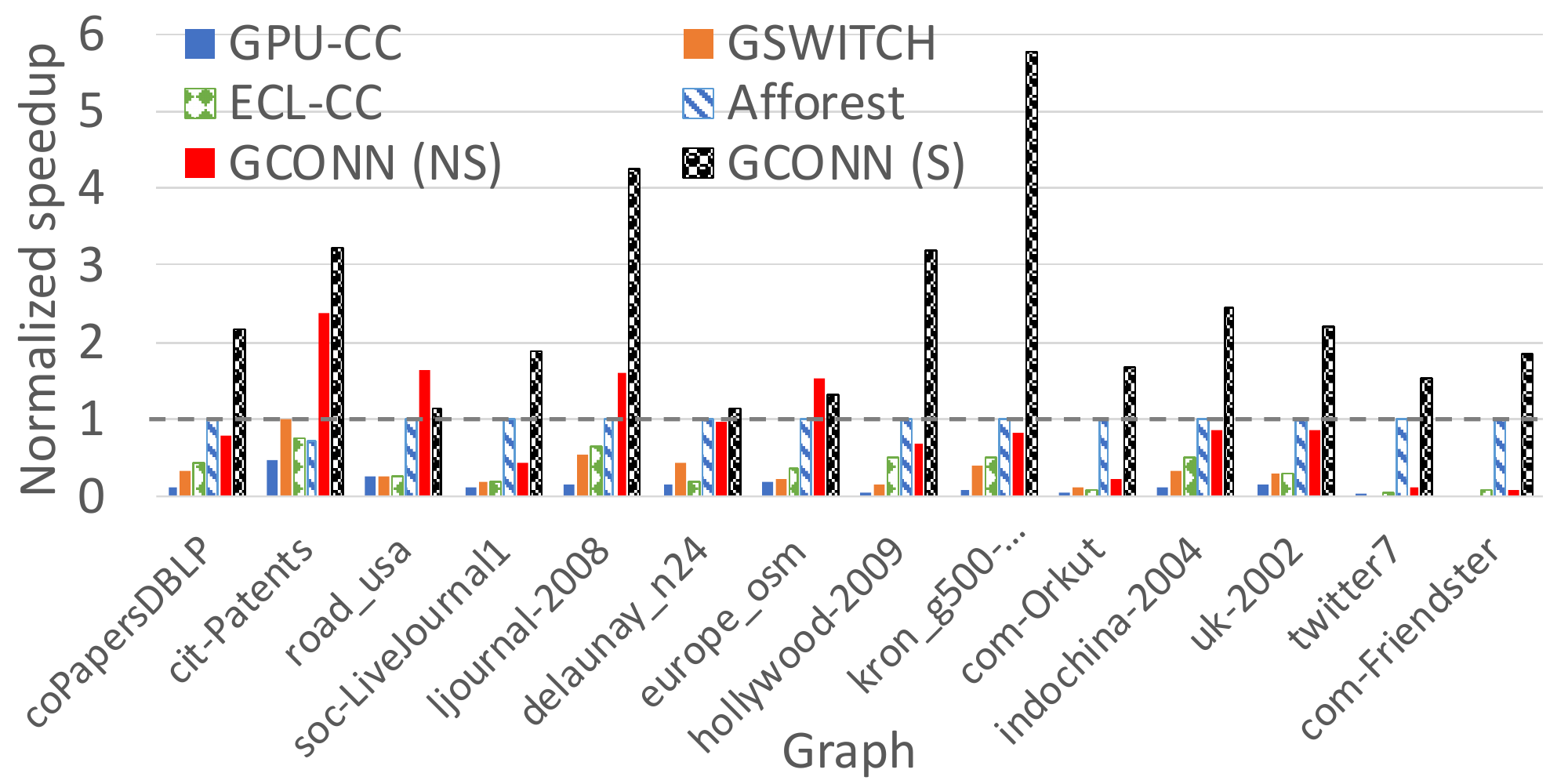}
        \vspace{-2em}
        \caption{The normalized speedup over \gpucc{} of current state-of-the-art implementations/libraries and \framework{} (with and without sampling).
		\CONV{GCONN(NS)} and \CONV{GCONN(S)} show the normalized speedup without and with sampling, respectively.}
        \label{fig:state}
\end{figure}

This section compares our implementations with current state-of-the-art connectivity implementations on GPUs, which are shown in
the last group of rows in Table~\ref{tab:overall_gpu}.
Figure~\ref{fig:state} presents the normalized execution time over the fastest current state-of-the-art implementation for each graph with our fastest variant without and with sampling.
Note that we did not report results for \CONV{Groute}~\cite{Ben-Nun2017}, \CONV{Gunrock}~\cite{Wang2016}, and \CONV{IrGL}~\cite{Pai2016}
as they are outperformed by \ORIeclcc{} \cite{Jaiganesh2018}, which we compare against.
We also tried the strategy by Cong and Muzio~\cite{Cong2014}, but it never gave the best performance.

\myparagraph{\gpucc{}~\cite{Soman} and \gswitch{}~\cite{Meng2019}} \gpucc{} adopts a classic SV algorithm, which turns out to be slow as shown in Figure~\ref{fig:state}.
Our fastest implementations without sampling and with sampling are 7.06x and 26.68x  faster \avg{} than \gpucc{}, respectively.
\gswitch{} also applies a classic SV algorithm, but outperforms \gpucc{} due to several additional optimizations.
Our fastest implementations without sampling and with sampling are 3.24x and 8.88x faster \avg{} than \gswitch{}.

\myparagraph{\ORIeclcc{}~\cite{Jaiganesh2018}}
\ORIeclcc{} is faster than both \gpucc{} and \gswitch{} due to using a more efficient algorithm based on union-find.
As shown in Figure~\ref{fig:state}, our fastest implementations without sampling and with sampling are 2.78x and 10.15x faster \avg{} than \ORIeclcc{}, respectively,
mainly because we use a more efficient load-balancing strategy, and more efficient find and compress rules.

We also implemented the finish algorithm \eclcc{} that is used in \ORIeclcc{}.
As shown in Table~\ref{tab:overall_gpu}, \eclcc{} without sampling is 2.30x faster on average than \ORIeclcc{}.
The speedup comes from using a different load-balancing strategy; in \ORIeclcc{}, one thread handles all edges of a vertex with degree at most 16,
which can lead to significant load-imbalance. In contrast, \framework{} enables those edges to be processed in a load-balanced fashion using a variation of a strategy by Merrill et al.~\cite{Merrill2012} that we designed.
Without sampling and with sampling, \eclcc{} is 1.19x and 1.04x slower \avg{} than our fastest variants, respectively.

\myparagraph{\ORIafforest{}~\cite{Sutton2018}} \ORIafforest{} also implements a root-based algorithm with \koutsample{}.
As shown in Figure~\ref{fig:state}, our fastest implementations without sampling are 1\% slower \avg{} than \ORIafforest{} with sampling
due to the fact that many edges are not inspected in \ORIafforest{}. 
Our fastest implementations with sampling are 2.51x faster \avg{} because
variants using \hbsample{} sometimes outperform those using \koutsample{},
and the finish algorithm used in \CONV{Afforest} is slower than those used in our fastest variants. 
We also implemented \afforest{}, the finish algorithm used in \ORIafforest{}, in \framework{}.
We found that compared to other finish algorithms in the no sampling setting, \afforest{} is 12.6x slower \avg{}  than our fastest finish algorithm.
The main reason seems to be due to some union-find algorithms, such as \unionasync{} and \unionremcas{}, handling path compression more efficiently than the method used in \afforest{}.

\subsection{Incremental Parallel Graph Connectivity}\label{sec:streaming_gpu}

In this section, we evaluate incremental connectivity algorithms on GPUs.
We achieve a raw speedup of 2,482x over \CONV{EvoGraph}~\cite{Sengupta17} which is the fastest current state-of-the-art streaming connectivity implementation. 
Furthermore, the memory bandwidth-normalized speedup is 794x over \CONV{EvoGraph}.
Unfortunately, we are not able to directly compare our streaming connectivity
implementations with existing implementations on the same machine,
although we provide indirect comparisons based on the numbers provided
by the authors in their paper.\footnote{We contacted the authors for their code, but were unable to obtain it.}
\CONV{cuSTINGER}~\cite{Green16} and \CONV{Hornet}~\cite{busato2018hornet} are GPU frameworks for streaming graphs, but their code does not contain streaming connectivity algorithms. 

We conduct two types of streaming experiments. In the first type of experiment, we generate a stream of edge updates from the input graphs in Table~\ref{table:sizes}.
The second type of experiment models real-world graph streams using synthetic graph generators. Specifically, we use the RMAT generator~\cite{ChakrabartiZF04} with parameters $(a,b,c) = (0.5,0.1,0.1)$ and the Barabasi-Albert (BA) generator~\cite{scalefree}.
For both generators, we use $n = 2^{27}$ and $m=10n$.
The edges in a batch are given in COO format, and are unsorted.

\begin{table*}[!t]
  \footnotesize
        \setlength{\tabcolsep}{0.5pt}
\centering
\begin{tabular}[t]{ @{} l | c | c | c | c | c | c | c | c | c | c | c | c | c | c | c | c}
  \toprule
  \multicolumn{1}{l}{\bf \begin{tabular}[c]{@{}c@{}}Algorithm\end{tabular}}
              & \multicolumn{1}{c}{\bf \begin{tabular}[c]{@{}c@{}}coPapers\\DBLP\end{tabular}} & \multicolumn{1}{c}{\bf \begin{tabular}[c]{@{}c@{}}cit-\\Patents\end{tabular}}
& \multicolumn{1}{c}{\bf road\_usa} & \multicolumn{1}{c}{\bf \begin{tabular}[c]{@{}c@{}}soc-Live\\Journal1\end{tabular}} &  \multicolumn{1}{c}{\bf \begin{tabular}[c]{@{}c@{}}ljournal\\-2008\end{tabular}} &
\multicolumn{1}{c}{\bf  \begin{tabular}[c]{@{}c@{}}delaunay\\\_n24\end{tabular}} & \multicolumn{1}{c}{\bf \begin{tabular}[c]{@{}c@{}}europe\\\_osm\end{tabular}} &
          \multicolumn{1}{c}{\bf \begin{tabular}[c]{@{}c@{}}hollywood\\-2009\end{tabular}}  & \multicolumn{1}{c}{\bf \begin{tabular}[c]{@{}c@{}}kron\_g500\\-logn21\end{tabular}}  &
                  \multicolumn{1}{c}{\bf \begin{tabular}[c]{@{}c@{}}com-\\Orkut\end{tabular}} & \multicolumn{1}{c}{\bf \begin{tabular}[c]{@{}c@{}}indochina\\-2004\end{tabular}} & \multicolumn{1}{c}{\bf uk-2002}
& \multicolumn{1}{c}{\bf twitter7} & \multicolumn{1}{c}{\bf \begin{tabular}[c]{@{}c@{}}com-\\Friendster\end{tabular}} &  \multicolumn{1}{c}{\bf RMAT} & \multicolumn{1}{c}{\bf BA} \\
  \midrule

\unionearly{}&		8.18e09&	6.08e09&	5.44e09&	5.19e09&	3.78e09&	4.87e09&	\G{6.83e09}&	4.65e09&	4.37e09&	5.58e09&	2.39e09&	3.42e09&	3.19e09&	1.78e09&	3.67e09&	7.50e08 \\
\unionhook{}&		2.66e10&	7.51e09&	5.69e09&	1.17e10&	9.58e09&	8.78e09&	4.75e09&	3.52e10&	2.65e10&	4.26e10&	1.29e10&	1.29e10&	1.18e10&	5.54e09&	9.95e09&	4.81e09 \\
\unionasync{}&		\G{3.17e10}&	\G{1.27e10}&	\G{7.35e09}&	\G{1.72e10}&	\G{1.31e10}&	\G{1.24e10}&	6.74e09&	\G{3.93e10}&	\G{3.25e10}&	\G{4.66e10}&	\G{1.59e10}&	\G{1.65e10}&	\G{1.51e10}&	6.08e09&	\G{1.17e10}&	5.03e09 \\
\unionremcas{}&		1.05e10&	6.50e09&	4.46e09&	6.20e09&	4.00e09&	5.70e09&	6.05e09&	7.64e09&	1.03e10&	1.16e10&	1.98e09&	3.27e09&	3.80e09&	3.14e09&	5.60e09&	9.61e08 \\
\unionremlock{}&	3.25e09&	5.45e09&	3.48e09&	1.96e09&	4.97e08&	5.50e09&	4.44e09&	6.20e09&	1.94e10&	4.49e10&	4.14e07&	1.55e08&	5.46e08&	\G{7.23e09}&	1.11e10&	4.17e09 \\
\jayanti{}&		2.78e10&	8.70e09&	5.29e09&	1.33e10&	9.39e09&	8.63e09&	4.68e09&	3.32e10&	2.52e10&	2.94e10&	1.14e10&	1.21e10&	1.41e10&	5.85e09&	8.14e09&	\G{5.70e09} \\
\liutarjan{} &		1.39e10&	2.50e09&	2.92e09&	7.41e09&	6.99e09&	4.62e09&	2.63e09&	1.42e10&	7.38e09&	1.18e10&	8.85e09&	9.56e09&	4.74e09&	2.08e09&	3.05e09&	2.28e09 \\
\shiloachvishkin{}&	9.42e09&	2.09e09&	2.97e09&	5.39e09&	4.27e09&	3.12e09&	2.02e09&	4.84e09&	3.90e09&	3.79e09&	5.90e09&	7.90e09&	3.22e09&	1.37e09&	2.50e09&	2.36e09 \\

\end{tabular}
\caption{Throughput achieved by incremental connectivity
  algorithms in \framework{} on a V100 GPU machine when all of the
  edges in the graph are treated as a single batch of updates.  For
  each graph, the highest throughput is shown in green. }
\label{tab:streaming_gpu}
\end{table*}

\myparagraph{Throughput}
We evaluate the throughput of our incremental connectivity implementations  on all of the input graphs in Table~\ref{tab:overall_gpu},
and the two large synthetic graphs generated from the RMAT and BA generators.
Table~\ref{tab:streaming_gpu} reports the streaming throughput
achieved by the fastest variant of each algorithm for each input when all of the edges are treated as a single batch of updates.

As shown in Table~\ref{tab:streaming_gpu}, the \unionasync{} algorithm
usually achieves the highest throughput.
Recall that for static parallel connectivity, \unionremcas{} is 1.02x slower \avg{} than \unionasync{}, but
here \unionremcas{} is 3.41x slower \avg{} than \unionasync{}. 
In the incremental setting, when accessing an element of the $\mathit{labels}$ array,
the algorithm must check whether the element has been initialized,
which incurs an additional overhead:
for instance, \ensuremath{\mathsf{Union\mhyphen}} \ensuremath{\mathsf{Async}} when used in the incremental setting is 1.73x slower \avg{} than when it is used in the static setting.
Compared to \unionasync{}, in \unionremcas{}, the $\mathit{labels}$ array is accessed much more often, which decreases the throughput. 

The other min-based algorithms are much slower than the union-find algorithms because the other min-based algorithms inspect more edges.
In particular, \liutarjan{} is 2.86x slower \avg{} than \unionasync{}, and
\shiloachvishkin{} is 4.87x slower \avg{} than \unionasync{}. 
 
\begin{figure}[!t]
	\centering
	\includegraphics[width=\linewidth]{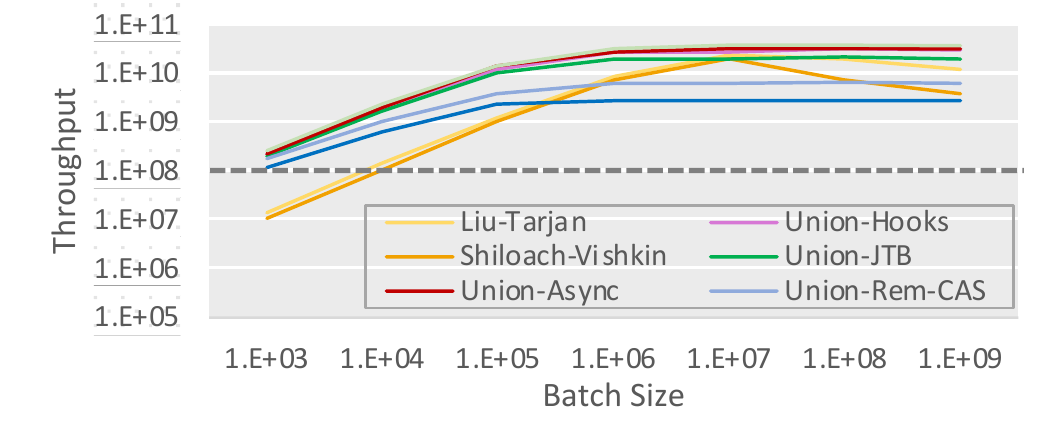}
	\vspace{-2em}
	\caption{Throughput vs. batch size for \emph{com-Orkut}.}
	\label{fig:batch_size_vs_throughput_gpu}
\end{figure}

\begin{figure}[!t]
	\centering
	\includegraphics[width=\linewidth]{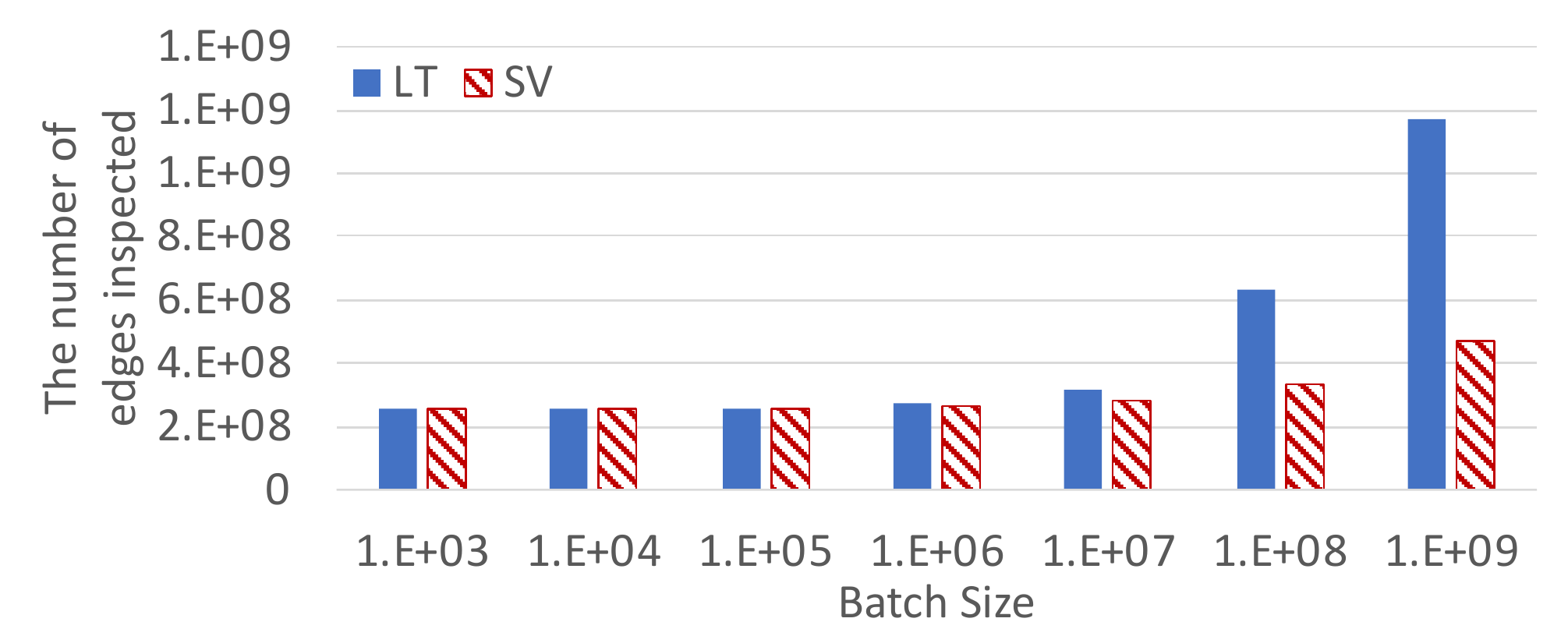}
	\vspace{-2em}
	\caption{The number of edges inspected for each batch size for \liutarjan{} \fullver{(the \ensuremath{\mathsf{PRS}} variant)}{} and \shiloachvishkin{} on \emph{com-Orkut}.}
	\label{fig:edge_inspect}
\end{figure}

\myparagraph{Throughput vs. Batch Size}
Figure~\ref{fig:batch_size_vs_throughput_gpu} shows the throughput
of the fastest variant
for each algorithm with respect to different batch sizes.
When the batch size is small, the performance significantly degrades mainly due to the GPU kernel launch overhead:
the time to launch the GPU kernel takes tens of times longer than the time to run it for a small batch size.
For small batch sizes, the \liutarjan{} and \shiloachvishkin{} algorithms are much slower because for each batch, the GPU kernels are launched multiple times to process the current batch until the $\mathit{labels}$ array  converges.

One interesting point is that  \liutarjan{} and \shiloachvishkin{} become slower when the batch size is larger than $10^{7}$ due to the increase in the number of edge inspections.
Figure~\ref{fig:edge_inspect} shows the number of edges being inspected for different batch sizes for one variant \fullver{(\ensuremath{\mathsf{PRS}}, described in Appendix~\ref{apx:lt}) of}{of}  \liutarjan{} as well as  \shiloachvishkin{}  on \emph{com-Orkut}.
The number of edges examined  increases significantly for a batch size  larger than $10^{7}$
because in those algorithms,
all edges in a batch need to be inspected until the batch converges.
We observed a similar trend in the other graphs for all variants of \liutarjan{} as well as \shiloachvishkin{}.

\begin{figure}[!t]
	\centering
	\includegraphics[width=\linewidth]{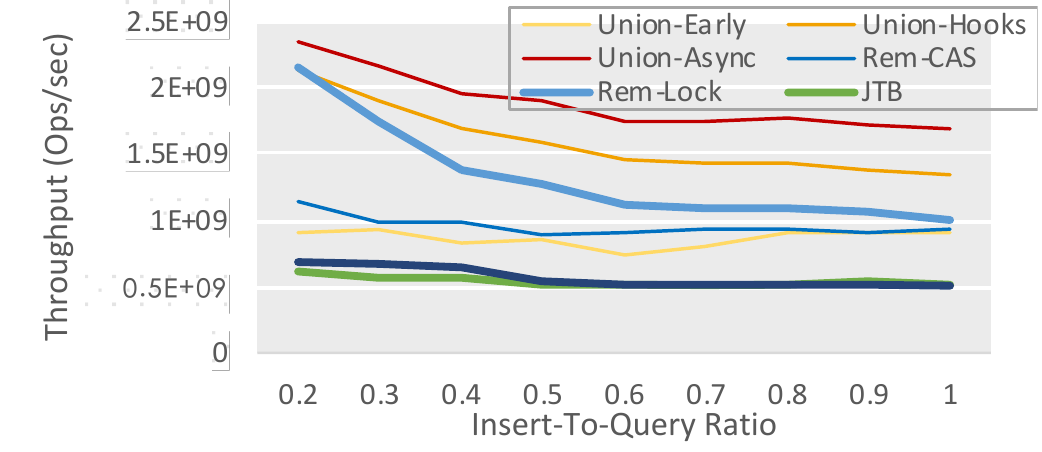}
	\vspace{-2em}
	\caption{Throughput of the fastest variant (when the below ratio is 1) of each algorithm on \emph{europe\_osm},
	plotted for different ratios of insertions to queries.}
	\label{fig:query}
\end{figure}

\myparagraph{Mixed Inserts and Queries}
We evaluate the performance of the variants used in Figure~\ref{fig:batch_size_vs_throughput_gpu}
to study how the ratio of insertions to queries affects the throughput.
For a ratio of insertions to queries of $x$, we generate $1/x$ queries with random vertex pairs per insert,
and shuffle them with the original edges (which are insertions) that are also randomly permuted
to prevent data locality from affecting performance. 

Figure~\ref{fig:query} shows the throughput with different insert-to-query ratios on the \emph{europe\_osm} graph.
As the ratio decreases, throughput also increases
because the path to the root vertex in the $\mathit{labels}$ array is compressed,
which speeds up the processing of the remaining insertions and queries.
Note that for very well connected graphs (e.g., \emph{com-Orkut}), the throughput with respect to different ratios is quite stable,
because even when the ratio is 1, the labels have converged after processing a small subset of insertions.

\subsection{Evaluation on Titan Xp (Pascal)}

\begin{table}[]
\center
\footnotesize
\begin{tabular}{|c|c|c|c|}
\hline
\multirow{2}{*}{Algorithm} & \multicolumn{2}{c|}  {Average slowdown against V100} & \multirow{2}{*}{\begin{tabular}[c]{@{}c@{}}Average speedup\\ over state-of-the-art\end{tabular}} \\ \cline{2-3}
                           & Across all variants   & Fastest variants   &                                                                                          \\ \hline
Static conn.        & 1.62                  & 1.71               & 2.38                                                                                     \\ \hline
Span. forest            & 1.59                  & 1.60               & --                                                                                       \\ \hline
Streaming conn.     & 1.79                  & 2.25               & --                                                                                       \\ \hline
\end{tabular}
\caption{ Summary of results on the Titan Xp machine.
}
\label{tab:pascal}
\end{table}

This section presents our evaluation of \framework{} on the Titan Xp (Pascal) GPU.
Table~\ref{tab:pascal} shows the average slowdown against our
evaluation on the V100 GPU,
as well as the average of the speedup over the fastest one among
\gpucc{}, \gswitch{}, \ORIeclcc{}, and \ORIafforest{} also
run on the Titan Xp GPU.
As shown in Table~\ref{tab:pascal}, the trends for both GPU machines are very similar, and except for incremental connectivity,
the average slowdown is close to the ratio of the bandwidth of the V100 to that of Titan Xp ($900/547.6=1.6$).
For incremental connectivity, accessing the $\mathit{labels}$ array requires spin-locks, which prevents the algorithms from fully saturating the
memory bandwidth.
We note that we had to modify \unionremlock{} to avoid
deadlocks~\cite{o2011parallel,jason2010cuda} for the Pascal machine in
which threads in a warp are executed in 
lock-step~\cite{jia2018dissecting}.

\subsection{Performance analysis on CPUs vs. GPUs}\label{sec:gpu_vs_gpu}

\begin{figure}[!t]
        \centering
        \includegraphics[width=\linewidth]{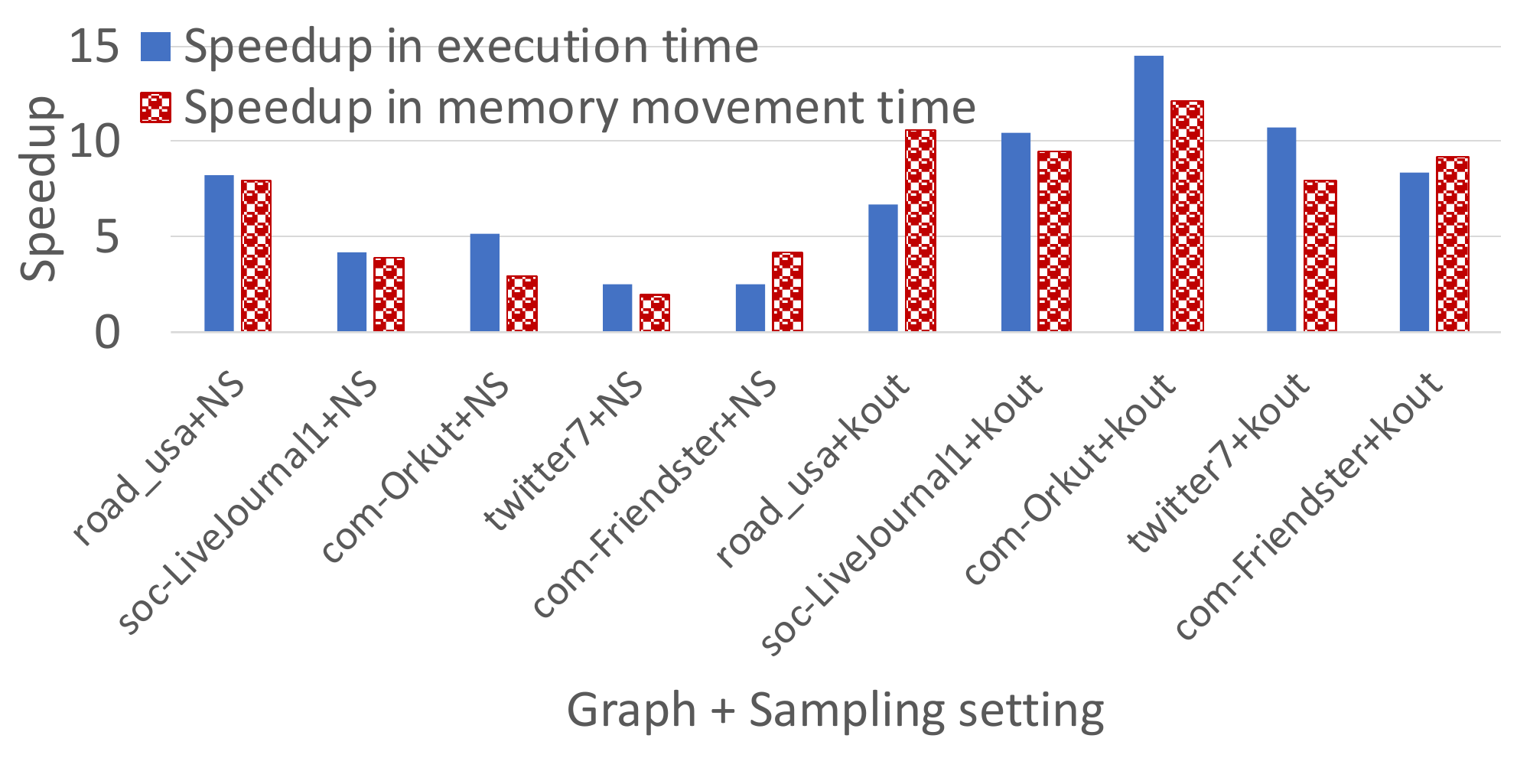}
        \vspace{-1em}
        \caption{The performance comparison between the fastest variant on the GPU vs. CPU
         for each graph in
        the no-sampling setting (\emph{+NS}) and with \koutsample{} (\emph{+kout}).
	}
        \label{fig:gpu_vs_cpu}
\end{figure}

In this section, we compare the performance of \framework{}~\cite{CONNECTIT} on the V100 with the performance of \connectit{}.
\connectit{}'s experiments were performed on a Dell PowerEdge R930 with
4 $\times$ 2.4GHz Intel 18-core E7-8864 x4 Xeon professors, a 45MB L3
Cache and a memory bandwidth of 85 GB/sec.

We compare CPU and GPU performance with the five graphs used in \connectit{} that can fit in the GPU memory~\cite{CONNECTIT}.
To understand the relation between the execution time and the total data movement from/to memory,
Figure~\ref{fig:gpu_vs_cpu} shows the speedup in execution time achieved by \framework{} over \connectit{} (blue bars),
and the  speedup in time for the data movement from/to memory at peak memory bandwidth using the GPU compared to the CPU (red bars).
Our GPU provides $900/85=10.59$ times higher memory bandwidth compared to the CPU, but the CPU L3 cache is much larger than the GPU L2 cache.
Hence, there can be more memory transactions on the GPU, depending on the graph.
As shown in Figure~\ref{fig:gpu_vs_cpu}, there is a strong correlation between the execution time and the time for data movement (a Pearson correlation coefficient of $0.854$).
Note that the first five pairs of bars are without sampling, and the rest of them are with \koutsample{}. For a fair comparison, we did not apply \hbsample{}
because the best variants in \connectit{} for these graphs always adopt \koutsample{}.
The best variant of \framework{} without \hbsample{} achieves 8.26--14.51x speedup over the best variant with \connectit{}.
For incremental algorithms,
when we treat all of the edges as one large batch,
\framework{} only achieves 1.85--13.36x speedup over \connectit{} 
due to the spin-locks on the $\mathit{labels}$ array being more expensive on GPUs than CPUs.

To obtain a rough estimate of the monetary cost savings for running the
connectivity algorithms on the GPU vs. the CPU, we compare two machines on
Amazon EC2: the \emph{p3.2xlarge} configuration, which is very similar
to our V100 setup, and the \emph{x1.16xlarge} configuration, which
provides a multicore similar to the Dell PowerEdge R930.  On-demand
pricing of the \emph{p3.2xlarge} and \emph{x1.16xlarge} instances is
\$3.06 and \$6.669 per hour, respectively.  The fastest variant of
\framework{} is 12.02x faster on average than that of \connectit{}.
Hence, for our input graphs, we can expect to save roughly a factor of
$(6.669/3.06) \times 12.02=26.2$ in costs by using a similar GPU compared to a
similar CPU.

\subsection{Takeaways and Guidelines}
\label{sec:tune}
Based on our experimental study, we found that variants of union-find that have not been studied in prior work on GPU connectivity performed the best.  
We discovered that the sampling phase mostly dominates the execution time,
which indicates that reducing the overhead for the sampling phase is crucial for high performance.
We found that many of our static connectivity algorithms can be extended to support spanning forest and incremental connectivity, and can achieve high performance as well.
Finally, we found GPUs to be a cost-efficient option for connectivity algorithms compared to CPUs, as long as the graph can fit in the GPU memory.

The fastest implementation is dependent on a given input graph, and as \framework{} supports several hundred variants, trying all variants to find the fastest one can be overwhelming. We provide some guidelines below on how to choose an implementation with high performance.

First, a practitioner can simply use the best variant overall based on
our experimental evaluation, which is to use \unionasync{}
or \unionremcas{} combined with \koutsample{}. For our input graphs,
using this approach achieves performance that is within 20.8\% on
average of the performance of the fastest implementation for each
graph.

\begin{figure}[!t]
        \centering
        \includegraphics[width=\linewidth]{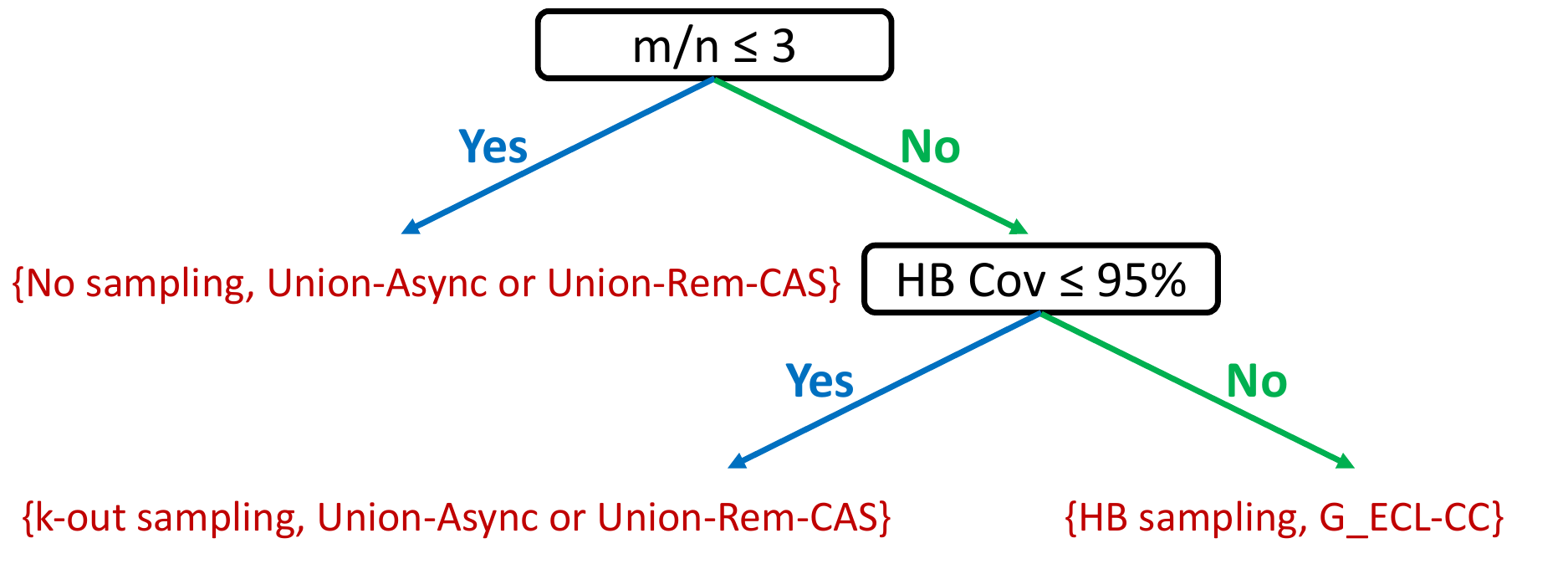}
        \vspace{-1em}
        \caption{A decision tree for the sophisticated dynamic mechanism.}
        \label{fig:decision_tree}
\end{figure}

A more sophisticated dynamic mechanism for selecting sampling methods
for a fixed finish method is as follows. First, we observed that if the average
degree in the graph is small enough, most edges are inspected in the
sampling phase, and an additional overhead to access the edge array
multiple times is incurred. Therefore, we found that when the average
degree is small, sampling is not a worthwhile optimization.
Second, we observed that one
does not need to consider BFS sampling, as it is always outperformed
by either \koutsample{} or hook-based
search (HB) sampling.
From Table~\ref{tab:sample_cover}, we see that when {\bf HB Cov} (the percentage of vertices in the largest connected
component after HB sampling) is high enough, HB sampling is always the
best strategy. As shown in Figure~\ref{fig:sample_comp}, in this case, HB sampling samples
a minimal number of edges. 
A rough lower bound of {\bf HB Cov} can be obtained by applying the first step of HB for a few edges, which is
inexpensive, and then using this value to choose between HB
or \koutsample{}.

Our study also provides insights on choosing finish methods that complement a particular sampling method. First, we found that the fastest finish method is always one of $\{$\unionasync{}, \ensuremath{\mathsf{Union\mhyphen Rem\mhyphen}} \ensuremath{\mathsf{CAS}}, or \eclcc{}$\}$.
In the no-sampling setting, either \ensuremath{\mathsf{Union\mhyphen}} \ensuremath{\mathsf{Async}} or \unionremcas{} is the fastest.
\eclcc{} performs fewer on-the-fly compressions, which normally leads to more path traversal (and thus more cache misses). 
On the other hand, when \hbsample{} is used, which implies the input graph is very well connected, \eclcc{} is recommended as only a very few edges are inspected by the union-find algorithms.
In this case, \eclcc{} has the lowest overhead as it does not perform much on-the-fly compression, which requires performing additional writes. 
On our input graphs, using this strategy for choosing the sampling and finish methods gives
a slowdown of only 9.4\% on average
compared to using the fastest implementation for each graph.

Figure~\ref{fig:decision_tree} presents a decision tree for the strategy above. {\bf HB Cov} is the lower bound mentioned above, and \unionasync{} can be substituted with \unionremcas{} as their performance is similar.

\section{Conclusion}\label{sec:future}

We have designed the \framework{} framework, which supports several hundred
efficient implementations on GPUs for static connectivity, spanning
forest, and incremental connectivity. To the best of our knowledge,
this paper provides the most comprehensive study of different
variants of connectivity algorithms on GPUs. Extensive evaluations
show that the best connectivity implementations in \framework{} significantly outperform other state-of-the-art libraries and
implementations.
For future work, we are interested in extending our framework to the multi-GPU or distributed memory settings.

\section*{Acknowledgements}
We thank the reviewers of this paper for their helpful feedback.
This research was supported by DOE Early Career Award \#DE-SC0018947,
NSF CAREER Award \#CCF-1845763, Google Faculty Research Award, DARPA
SDH Award \#HR0011-18-3-0007, and Applications Driving Architectures
(ADA) Research Center, a JUMP Center co-sponsored by SRC and DARPA.

\bibliographystyle{ACM-Reference-Format}
\bibliography{references.bib}

\fullver{\begin{appendix}


\section{\framework{} Pseudocode}\label{apx:pseudocode}

\algblock{ParFor}{EndParFor}
\algnewcommand\algorithmicparfor{\textbf{parfor}}
\algnewcommand\algorithmicpardo{\textbf{do}}
\algnewcommand\algorithmicendparfor{}
\algrenewtext{ParFor}[1]{\algorithmicparfor\ #1\ \algorithmicpardo}
\algrenewtext{EndParFor}{\algorithmicendparfor}
\algtext*{EndParFor}{}

\subsection{Sampling Schemes}\label{apx:subsec:sampling_pseudocode}
Here we provide the pseudocode for our sampling methods in
\framework{}.
Algorithm~\ref{alg:kout_sample} provides pseudocode for the
\koutsample{} method, with parameter $k$. We set $k=2$ in
our experiments.
Algorithm~\ref{alg:hb_sample} provides pseudocode for the
\hbsample{} method, with parameter $N$.
We set $N=4$ in our experiments.
Algorithm~\ref{alg:bfs_sample} provides pseudocode for the
\bfssample{} method. The BFS starts from the
vertex with the highest degree in the graph.

\begin{algorithm}[!t]
\caption{\koutsample{}} \label{alg:kout_sample}
\small
\begin{algorithmic}[1]
\Procedure{$k$-out-Sampling}{$G(V, E), \codevar{labels}, k$}
  \State $\codevar{edges} \gets$ $\{$ First edge from each vertex $\} \cup \{$Sample $k-1$ edges uniformly at random from each vertex$\}$
  \State $\textsc{UnionFind}(\codevar{edges}, \codevar{labels})$
  \State Fully compress the components array in parallel.\label{line:koutcompress}
  \State  \algorithmicreturn{} $\codevar{labels}$
\EndProcedure
\end{algorithmic}
\end{algorithm}

\begin{algorithm}[!t]
\caption{\hbsample{}} \label{alg:hb_sample}
\small
\begin{algorithmic}[1]
\Procedure{HB-Sampling}{$G(V, E), \codevar{labels}, N$}
  \State $\codevar{labels} \gets \{\codevar{labels}[v] \gets \min(v, w) \ |\ v \in V\}$ \Comment{$w$ is the smallest neighbor of $v$} 
  \State $\codevar{edges} \gets$ $\{$ First $N$ edges from each vertex $v$ such that $\codevar{labels[v]} \neq v$ $\}$
  \State $\textsc{UnionFind}(\codevar{edges}, \codevar{labels})$
  \State Fully compress the components array in parallel.\label{line:koutcompress}
  \State  \algorithmicreturn{} $\codevar{labels}$
\EndProcedure
\end{algorithmic}
\end{algorithm}


\begin{algorithm}[!t]
  \caption{\bfssample{}} \label{alg:bfs_sample}
\small
\begin{algorithmic}[1]
\Procedure{BFS-Sampling}{$G(V, E), \codevar{labels}$}
    \State $s \gets \argmax_{v \in V_{s}} d(v)$ \Comment{$V_{s}$ is a sampled subset of $V$} 
    \State $\codevar{labels} \gets \textsc{LabelSpreadingBFS}(G, s)$
    \State $\codevar{freq} \gets \textsc{IdentifyFrequent}(\codevar{labels})$
      \State  \algorithmicreturn{} $\codevar{labels}$
\EndProcedure
\end{algorithmic}
\end{algorithm}

\begin{algorithm*}[!t]
  \caption{\ufmetaalgorithm{}} \label{alg:union_find_meta}
\begin{algorithmic}[1]
\State $\mathsf{FindOption} = \{$ \Comment{Shared by union-find algorithms}
  \State \hspace{1em}$\findnaive{}$, \Comment{No path compression}
  \State \hspace{1em}$\findsplit{}$, \Comment{Path splitting}
  \State \hspace{1em}$\twotrysplit{}$, \Comment{Two-try splitting (only for $\mathsf{\jayanti{}}$)}
  \State \hspace{1em}$\findhalve{}$, \Comment{Path halving}
  \State \hspace{1em}$\findcompress{}$, \Comment{Full path compression}
\State $\}$
\State $\mathsf{SpliceOption} = \{$ \Comment{Used by Rem's algorithms when path has not yet reached a root}
  \State \hspace{1em}$\splitone{}$, \Comment{Performs one path split}
  \State \hspace{1em}$\halveone{}$, \Comment{Performs one path halve}
  \State \hspace{1em}$\splice{}$, \Comment{Performs a splice operation}
\State $\}$
\State $\mathsf{UnionOption} = \{$
  \State \hspace{1em}$\mathsf{\unionasync{}}$, \Comment{Asynchronous union-find}
  \State \hspace{1em}$\mathsf{\unionhook{}}$, \Comment{Asynchronous hook-based union-find}
  \State \hspace{1em}$\mathsf{\unionearly{}}$, \Comment{Union with early-optimization}
  \State \hspace{1em}$\mathsf{\unionremlock{}}$, \Comment{Lock-based Rem's algorithm}
  \State \hspace{1em}$\mathsf{\unionremcas{}}$, \Comment{CAS-based Rem's algorithm}
  \State \hspace{1em}$\mathsf{\jayanti{}}$, \Comment{Randomized algorithm by Jayanti, Tarjan, and Boix{-}Adser{\`{a}}}
  \State \hspace{1em}$\mathsf{\eclcc{}}$, \Comment{The union-find algorithm used in \ORIeclcc{}}
  \State \hspace{1em}$\mathsf{\afforest{}}$, \Comment{The union-find algorithm used in \ORIafforest{}}
\State $\}$
\Procedure{Connectivity}{$G$, $\codevar{labels}$,
$\codevar{L_{\max}}$, $\mathsf{UnionOption}$, $\mathsf{FindOption}$,
  $\mathsf{SpliceOption}$}
\State $\mathcal{U} = \mathsf{Union\mhyphen Find}(\mathsf{UnionOption}$, $\mathsf{FindOption}$, $\mathsf{SpliceOption})$ 
\ParFor {$\{u\ |\ \codevar{labels}[u] \neq \codevar{L_{\max}}\}$}
  \ParFor {$\{(u, v) \in d^{+}(u)\}$}
    \State $\mathcal{U}.\mathsf{Union}(u, v, \codevar{labels})$
  \EndParFor
\EndParFor
\EndProcedure
\end{algorithmic}
\end{algorithm*}

\begin{algorithm}[!t]
  \caption{Find Algorithms} \label{alg:find_options}
\small
\begin{algorithmic}[1]
\Procedure{FindNaive}{$u, P$}
  \State $v \gets u$
  \While {$v \neq P[v]^{*}$}
    \State $v \gets P[v]$
  \EndWhile
  \State \algorithmicreturn{} $v$
\EndProcedure

\Procedure{FindCompress}{$u, P$}
  \State $r \gets u$
  \If {$P[r] = r$} \algorithmicreturn{} $r$
  \EndIf
  \While {$r \neq P[r]$}
    \State $r \gets P[r]$
  \EndWhile
  \While {$j \gets P[u] > r$}
  \State $P[u] \gets r$, $u \gets j$
  \EndWhile
  \State \algorithmicreturn{} $r$
\EndProcedure

\Procedure{\findsplit{}}{$u, P$}
  \State $v \gets P[u], w \gets P[v]$
  \While{$v \neq w^{*}$}
    \State $\textsc{\cas{}}(\&P[u], v, w)$
    \State $u \gets v$
  \EndWhile
  \State \algorithmicreturn{} $v$
\EndProcedure

\Procedure{\findhalve{}}{$u, P$}
  \State $v \gets P[u], w \gets P[v]$
  \While{$v \neq w^{*}$}
    \State $\textsc{\cas{}}(\&P[u], v, w)$
    \State $u \gets P[u]$
  \EndWhile
  \State \algorithmicreturn{} $v$
\EndProcedure
\end{algorithmic}
\end{algorithm}

\begin{algorithm}[!t]
  \caption{Splice Algorithms} \label{alg:splice_options}
\small
\begin{algorithmic}[1]
\Procedure{\splitone{}}{$u, x, P$}
  \State $v \gets P[r_u], w \gets P[v]^{*}$
  \If{$v \neq w$}
    \State $\textsc{\cas{}}(\&P[u], v, w)$
  \EndIf
  \State \algorithmicreturn{} $v$
\EndProcedure

\Procedure{\halveone{}}{$u, x, P$}
  \State $v \gets P[r_u], w \gets P[v]^{*}$
  \If{$v \neq w$}
    \State $\textsc{\cas{}}(\&P[u], v, w)$
  \EndIf
  \State \algorithmicreturn{} $w$
\EndProcedure

\Procedure{\splice{}}{$u, v, P$}
\State $p_u = P[u]^{*}$
  \State $\textsc{\cas{}}(\&P[u], p_u, P[v])$
  \State \algorithmicreturn{} $p_u$
\EndProcedure
\end{algorithmic}
\end{algorithm}

\subsection{Union-Find Algorithms}
In this section, we provide the pseudocode for our union-find
implementations.
Algorithm~\ref{alg:union_find_meta} shows our generic union-find
template algorithm, which takes a custom union operator, find
operator, and splice operator, and runs the resulting algorithm
combination. The splice operator is only valid for Rem's
algorithms. Algorithm~\ref{alg:find_options} provides pseudocode for
the different find implementations. Algorithm~\ref{alg:splice_options}
provides pseudocode for the different splice options.

Next, we provide pseudocode for each of the union-find implementations
in this paper, other than \jayanti{}, whose pseudocode is presented in
their paper~\cite{JayantiTB19}.

\myparagraph{Asynchronous Union-Find} The first
class of algorithms are inspired by a recent paper exploring
concurrent union-find implementations by Jayanti and
Tarjan~\cite{Jayanti2016}. We implement all of the variants from their
paper, as well as a full path compression technique (also considered
in~\cite{alistarh2019search}) which works better in practice in some
cases. We refer to this union-find algorithm as {$\bm{\unionasync{}}$}
(Algorithm~\ref{alg:unionasync}),
since it is the classic union-find algorithm directly adapted
for an asynchronous shared-memory setting. The algorithm links from
higher ID to lower ID vertices to avoid cycles, and only
performs links on roots. This algorithm can be combined with one of the following
implementations of the find operation: $\bm{\mathsf{FindNaive}}$,
which performs no compression during the operation,
$\bm{\mathsf{FindSplit}}$ and $\bm{\mathsf{FindHalve}}$, which perform
path-splitting and path-halving, respectively, and
$\bm{\mathsf{FindCompress}}$, which fully compresses the find path.
Jayanti and Tarjan show that this class of algorithms is linearizable
for a set of concurrent union and find operations (while they do not consider $\mathsf{FindCompress}$, it is relatively easy to show that it is linearizable).
Note that in Algorithms~\ref{alg:find_options}--\ref{alg:union_rem_cas}, we mark linearization points (which define a total order of the operations and are used to verify the correctness of algorithms~\cite{JayantiTB19}) with an asterisk (e.g., Line 4 in Algorithm~\ref{alg:find_options}).

\begin{algorithm}[!t]
  \caption{\unionasync{}} \label{alg:unionasync}
\small
\begin{algorithmic}[1]
  \State \textsc{Find} = \text{one of } $\{\mathsf{FindNaive}, \mathsf{\findsplit{}},$
  \State \hspace{7em} $\mathsf{\findhalve{}}, \mathsf{FindCompress}\}$
\Procedure{Union}{$u, v, P$}
\State $p_u \gets \textsc{Find}(u, P), p_v \gets \textsc{Find}(v, P)$
\While{$p_u \neq p_v$}\Comment{WLOG, let $p_u > p_v$}
\If{$p_u = P[p_u]$ and $\textsc{\cas{}}(\&P[u], p_u, p_v)$}
  \State \algorithmicreturn{}
  \EndIf
  \State $p_u \gets \textsc{Find}(u, P), p_v \gets \textsc{Find}(v, P)$
\EndWhile
\EndProcedure
\end{algorithmic}
\end{algorithm}

We also consider two similar variants of the \unionasync{} algorithm:
{\unionhook{}} (Algorithm~\ref{alg:union_nondeterministic}) and
{\unionearly{}} (Algorithm~\ref{alg:union_early}). $\bm{\unionhook{}}$ is closely
related to \unionasync{}, with the only difference between the
algorithms is that instead of performing a CAS directly on the array
storing the connectivity labeling, we perform a CAS on an auxiliary
$\codevar{hooks}$ array, and perform an uncontended write on the
$\codevar{parents}$ array. $\bm{\unionearly{}}$ is also similar to
\unionasync{}, except that the algorithm traverses the paths from both vertices together, and tries to eagerly check and
hook a vertex once it is a root.
The algorithm can optionally perform a find on the
endpoints of the edge after the union operation finishes, which has
the effect of compressing the find path. The linearizability proof for
\unionasync{} by Jayanti and Tarjan~\cite{Jayanti2016} applies almost
verbatim to \unionhook{} and \unionearly{}, and shows that both
algorithms are linearizable for a set of concurrent finds and unions.

\begin{algorithm}[!t]
  \caption{\unionhook{}} \label{alg:union_nondeterministic}
\small
\begin{algorithmic}[1]
\State \textsc{Find} = \text{one of } $\{\mathsf{FindNaive},\mathsf{\findsplit{}},$
\State \hspace{7em} $\mathsf{\findhalve{}}, \mathsf{FindCompress}\}$
\State $\codevar{H} \gets \{i \rightarrow \infty\ |\ i \in [0,\ldots,n-1] \}$ \Comment{Array of hooks}
\Procedure{Union}{$u, v, P$}
\State $p_u \gets \textsc{Find}(u, P), p_v \gets \textsc{Find}(v, P)$
\While{$p_u \neq p_v$} \Comment{WLOG, let $p_u > p_v$}
\If{$p_u = P[p_u]$ and $\textsc{\cas{}}(\&H[u], \infty, p_v)$}
  \State $P[p_u] \gets p_v$
  \State \algorithmicreturn{}
  \EndIf
  \State $p_u \gets \textsc{Find}(u, P), p_v \gets \textsc{Find}(v, P)$\Comment{WLOG, let $p_u > p_v$
  } 
\EndWhile
\EndProcedure
\end{algorithmic}
\end{algorithm}

\begin{algorithm}[!t]
  \caption{\unionearly{}} \label{alg:union_early}
\small
\begin{algorithmic}[1]
\State \textsc{Find} = \text{one of } $\{\mathsf{FindNaive},\mathsf{\findsplit{}},$
\State \hspace{7em}$\mathsf{\findhalve{}}, \mathsf{FindCompress}\}$
\Procedure{Union}{$u, v, P$}
\State $p_u \gets u, p_v \gets v$
\While{$p_u \neq p_v$} \Comment{WLOG, let $p_u > p_v$}
\If{$p_u = P[p_u]$ and $\textsc{\cas{}}(\&P[u], p_u, p_v)$}
    \State $\textsc{break}$
  \EndIf
  \State $z \gets P[p_u], w \gets P[z]$
  \State $\textsc{\cas{}}(\&P[p_u], z, w)$
  \State $p_u \gets w$
\EndWhile
\State $\textsc{Find}(u, P), \textsc{Find}(v, P)$ \Comment{Elided if the find-option is $\mathsf{FindNaive}$}
\EndProcedure
\end{algorithmic}
\end{algorithm}

\myparagraph{Concurrent Rem's Algorithms} We implement two concurrent versions of
Rem's algorithm, a lock-based version by Patwary et
al.~\cite{PatwaryRM12} ($\bm{\mathsf{UnionRemLock}}$,
shown in Algorithm~\ref{alg:union_rem_lock}) and a lock-free
compare-and-swap based implementation
($\bm{\mathsf{UnionRemCAS}}$, shown in Algorithm~\ref{alg:union_rem_cas}).
The lock-based version of Rem's
algorithm can be combined with the same rules for path compression
described for $\mathsf{Union}$ above, with the exception of
$\mathsf{FindCompress}$.
We also implemented another lock-based version of Rem's algorithm
to support NVIDIA GPUs before Volta generation.
The CAS-based version of Rem's algorithm
takes, in addition to the same full path compression strategies as
$\mathsf{UnionRemCAS}$, an extra \defn{splice} strategy, which
is used when a step of the union algorithm operates at a non-root
vertex. $\mathsf{UnionRemCAS}$ can be combined with one of the following
rules: $\bm{\mathsf{HalveAtomicOne}}$,
$\bm{\mathsf{SplitAtomicOne}}$, and $\bm{\mathsf{SpliceAtomic}}$.
The
first two rules perform a single path-halving or path-splitting, and
the second rule performs the splicing operation described in Rem's
algorithm. Rem's algorithm
is correct in a phase-concurrent setting, where unions and find
queries are separated by a barrier~\cite{CONNECTIT}.


\myparagraph{Randomized Two-Try Splitting}
Next, we incorporate a more sophisticated randomized algorithm by
Jayanti, Tarjan, and Boix{-}Adser{\`{a}}~\cite{JayantiTB19}, and
refer to this algorithm as $\bm{\jayanti{}}$. 
The algorithm either
performs finds naively, without using any path compression
($\bm{\mathsf{FindSimple}}$), or uses a strategy called
$\bm{\mathsf{FindTwoTrySplit}}$, which guarantees provably efficient bounds for
their algorithm, assuming a source of random bits. We
give a faithful implementation of the algorithm presented by Jayanti,
Jayanti, Tarjan, and Boix{-}Adser{\`{a}}~\cite{JayantiTB19}, and
refer to their paper for the pseudocode and proofs of correctness.
The \jayanti{} algorithm is linearizable, and only links roots.

\begin{algorithm}[!t]
  \caption{\unionremlock{}} \label{alg:union_rem_lock}
\small
\begin{algorithmic}[1]
 \State \textsc{Compress} = \text{one of } $\{\mathsf{FindNaive}, \mathsf{\findsplit{}}, \mathsf{\findhalve{}}\}$
\State \textsc{\splice{}} = \text{one of } $\{\mathsf{HalveAtomicOne},
\mathsf{SplitAtomicOne},$
\State \hspace{10em} $\mathsf{SpliceAtomic}\}$
\State $\codevar{L} \gets $ array of $n$ locks
\Procedure{Union}{$u, v, P$}
\State $r_u \gets u, r_v \gets v$
\While{$P[r_u] \neq P[r_v]^{*}$} \Comment{WLOG, let $P[r_u] > P[r_v]$}
  \If{$r_u = P[r_u]$}
    \State $L[r_u].\textsc{Lock}()$
    \State $p_v \gets P[r_v]$
    \If{$r_u = P[r_u]$ and $r_u > p_v^{*}$} 
      \State $P[r_u] \gets p_v$
    \EndIf
    \State $L[r_u].\textsc{Unlock}()$
    \State \algorithmicreturn{} $\mathsf{True}$
  \Else
    \State $r_u \gets \textsc{\splice{}}(r_u, r_v, P)$
  \EndIf
\EndWhile
\If {$\textsc{Compress} \neq \mathsf{FindNaive}$}
  \State $\textsc{Find}(u, P), \textsc{Find}(v, P)$
\EndIf
\State \algorithmicreturn{} $\mathsf{False}$
\EndProcedure
\end{algorithmic}
\end{algorithm}

\begin{algorithm}[!t]
  \caption{\unionremcas{}} \label{alg:union_rem_cas}
\small
\begin{algorithmic}[1]
  \State \textsc{Compress} = \text{one of } $\{\mathsf{FindNaive}, \mathsf{\findsplit{}}, \mathsf{\findhalve{}}\}$
\State \textsc{\splice{}} = \text{one of } $\{\mathsf{HalveAtomicOne},
\mathsf{SplitAtomicOne},$
\State \hspace{10em} $\mathsf{SpliceAtomic}\}$
\Procedure{Union}{$(u, v, P)$}
\State $r_u \gets u, r_v \gets v$
\While{$P[r_u] \neq P[r_v]^{*}$} \Comment{WLOG, let $P[r_u] > P[r_v]$}
\If{$r_u = P[r_u]$ and $\textproc{\cas{}}(\&P[r_u], r_u, P[r_v])^{*}$}
    \If {$\textsc{Compress} \neq \findnaive{}$}
      \State $\textsc{Compress}(u, P)$
      \State $\textsc{Compress}(v, P)$
    \EndIf
    \State \algorithmicreturn{} $\mathsf{True}$
  \Else
    \State $r_u \gets \textsc{\splice{}}(r_u, r_v, P)$
  \EndIf
\EndWhile
\State \algorithmicreturn{} $\mathsf{False}$
\EndProcedure
\end{algorithmic}
\end{algorithm}

\subsection{Shiloach-Vishkin Algorithm}
Algorithm~\ref{alg:shiloach_vishkin} shows the pseudocode for our
adaptation of the Shiloach-Vishkin algorithm.

\begin{algorithm}[!t]
  \caption{$\mathsf{Shiloach\mhyphen Vishkin}$}\label{alg:shiloach_vishkin}
\small
\begin{algorithmic}[1]
\Procedure{Connectivity}{$G$, $\codevar{labels}$, $\codevar{frequentid}$}
\State $\codevar{changed} \gets \mathsf{true}$
\State $\codevar{candidates} \gets \{ v \in V\ |\ \codevar{labels}[v] \neq \codevar{frequentid} \}$
\State $\codevar{prev\_labels} \gets \codevar{parents}$ \Comment{A copy}
\While {$\codevar{changed} = \mathsf{true}$}
  \State $\codevar{changed} = \mathsf{false}$
  \ParFor {$\{v \in \codevar{candidates}\}$}
    \ParFor {$\{(u, v) \in d^{+}(u)\}$}
      \State $p_u = \codevar{labels}[u], p_v = \codevar{labels}[v]$
      \State $l = \min(p_u, p_v), h = \max(p_u, p_v)$
      \If {$l \neq h$ and $h = \codevar{prev\_labels}[h]$}
        \State $\textsc{\writemin{}}(\&\codevar{labels}[h], l)$
        \State $\codevar{changed} \gets \mathsf{true}$
      \EndIf
    \EndParFor
  \EndParFor
  \ParFor {$\{v \in V\}$}
    \State $\textsc{FullShortcut}(v, \codevar{labels})$
    \State $\codevar{prev\_labels}[v] \gets \codevar{parents}[v]$
  \EndParFor
\EndWhile
\EndProcedure
\end{algorithmic}
\end{algorithm}

\subsection{Liu-Tarjan Algorithms} \label{apx:lt}
Finally, we describe the variants of the Liu-Tarjan framework
implemented in \framework{}.  Liu and Tarjan provide a framework for
simple and elegant concurrent connectivity algorithms based on several
simple rules which manipulate an array of parent pointers using edges
to transmit the connectivity information~\cite{LiuT19}. These
algorithms are not traditional concurrent algorithms, but actually
parallel algorithms designed for the Massively Parallel Computation
(\mpc{}) setting. We implement the framework proposed in their paper
as part of \framework{}, and also include several variants that can be
generated by their framework but were not considered in their paper
(only five variants were explored in their original paper). Their
framework ensures that the parent array is a \emph{minimum labeling},
where each vertex's parent is the minimum value among messages that it
has received.

Conceptually, each round of the algorithm processes all of the edges
and performs several rules on each edge. In each
round, each vertex receives a number of messages and updates its
parent at the end of the round to the minimum message it receives. Each
round performs a \emph{connect phase}, an \emph{update phase}, a
\emph{shortcut phase}, and possibly an \emph{alter phase}.
The \emph{connect phase} has several operations. A \connect{} sends
the endpoints of an edge to both vertices. A \parentconnect{} sends
the parents of the endpoints of an edge to both vertices. An
\extendedconnect{} sends the parent values of an edge to the edge
endpoints as well as the parents of the endpoints. Next, the framework
performs the \emph{update phase}, where there are two options. The
first option, \update{} sets the parent of every vertex to the minimum
received ID.  The second option, \rootupdate{} only updates the
parents of tree roots using an \update{}.  Next, the framework
performs the \emph{shortcut phase}, which calls \shortcut{} to
perform one step of path compression. The algorithm can
execute an optional \emph{alter phase}. \alter{} updates the endpoints
of an edge to be the current labels of the endpoints, which is
necessary when using \connect{}.

Below, we list all variants of the Liu-Tarjan algorithm that we
consider in this paper.

\begin{enumerate}[label=(\arabic*),itemsep=0pt]
  \item $\mathsf{CUSA}$: $\{\connect{}, \update{}, \shortcut{},\alter{}\}$
  \item $\mathsf{CRSA}$: $\{\connect{}, \rootupdate{}, \shortcut{},\alter{}\}$
  \item $\mathsf{PUSA}$: $\{\parentconnect{}, \update{}, \shortcut{}, \alter{}\}$
  \item $\mathsf{PRSA}$: $\{\parentconnect{}, \rootupdate{}, \shortcut{}, \alter{}\}$
  \item $\mathsf{PUS}$: $\{\parentconnect{}, \update{}, \shortcut{}\}$
  \item $\mathsf{PRS}$: $\{\parentconnect{}, \rootupdate{}, \shortcut{}\}$
  \item $\mathsf{EUSA}$: $\{\extendedconnect{}, \update{}, \shortcut{}, \alter{}\}$
  \item $\mathsf{EUS}$: $\{\extendedconnect{}, \update{}, \shortcut{}\}$

  \item $\mathsf{CUFA}$: $\{\connect{}, \update{}, \fullshortcut{},\alter{}\}$
  \item $\mathsf{CRFA}$: $\{\connect{}, \rootupdate{}, \fullshortcut{},\alter{}\}$
  \item $\mathsf{PUFA}$: $\{\parentconnect{}, \update{}, \fullshortcut{}, \alter{}\}$
  \item $\mathsf{PRFA}$: $\{\parentconnect{}, \rootupdate{}, \fullshortcut{}, \alter{}\}$
  \item $\mathsf{PUF}$: $\{\parentconnect{}, \update{}, \fullshortcut{}\}$
  \item $\mathsf{PRF}$: $\{\parentconnect{}, \rootupdate{}, \fullshortcut{}\}$
  \item $\mathsf{EUFA}$: $\{\extendedconnect{}, \update{}, \fullshortcut{}, \alter{}\}$
  \item $\mathsf{EUF}$: $\{\extendedconnect{}, \update{}, \fullshortcut{}\}$
\end{enumerate}

\end{appendix}
}{}


\end{document}